\documentclass[twocolumn,pra,superscriptaddress,notitlepage]{revtex4-1}
\usepackage{graphicx}
\usepackage{physics}
\usepackage{epsfig}
\usepackage{color}
\usepackage{xspace}
\usepackage{amsmath}
\usepackage{amssymb}
\usepackage{amsthm}
\usepackage{amsfonts}
\usepackage{ulem}
\usepackage[dvipsnames]{xcolor}
\usepackage{bm}
\usepackage{bbm}
\usepackage[breaklinks,colorlinks,linkcolor=blue,citecolor=blue,urlcolor=blue]{hyperref}
\usepackage{txfontsb}

\begin{document}
	
\global\long\def\id{\mathbbm{1}}
\global\long\def\ui{\mathbbm{i}}
\global\long\def\ud{\mathrm{d}}

\title{Proposal for realizing unpaired Weyl points in a three-dimensional periodically driven optical Raman lattice}

\author{Xiao-Dong Lin}
\affiliation{Hefei National Research Center for Physical Sciences at the Microscale and School of Physical Sciences, University of Science and Technology of China, Hefei 230026, China}
\affiliation{School of Physics and Institute for Quantum Science and Engineering, Huazhong University of Science and Technology, Wuhan 430074, China}
\author{Jinyi Zhang}
\affiliation{Hefei National Research Center for Physical Sciences at the Microscale and School of Physical Sciences, University of Science and Technology of China, Hefei 230026, China}
\affiliation{Shanghai Research Center for Quantum Sciences and CAS Center for Excellence in Quantum Information and Quantum Physics, University of Science and Technology of China, Shanghai 201315, China}
\affiliation{Hefei National Laboratory, Hefei 230088, China}
\author{Long Zhang}
\email{lzhangphys@hust.edu.cn}
\affiliation{School of Physics and Institute for Quantum Science and Engineering, Huazhong University of Science and Technology, Wuhan 430074, China}
\affiliation{Hefei National Laboratory, Hefei 230088, China}

\begin{abstract}
	In static lattice systems, the Nielsen-Ninomiya theorem enforces the pairing of Weyl points with opposite chiralities, which precludes the chiral magnetic effect (CME) in equilibrium. Periodic driving provides a viable route to circumvent this no-go constraint. Here, we propose a scheme to realize and control unpaired Weyl points using ultracold atoms in a three‑dimensional (3D) optical Raman lattice under continuous periodic driving. By engineering distinct relative symmetries between the lattice and multiple Raman potentials, the configuration generates an effective 3D spin-orbit coupling and yields a tunable topological‑insulator phase. Through adiabatic periodic modulation of this system, we show that eight Weyl points emerge in the quasienergy spectrum of the low‑energy sector, whose net chirality can be precisely tuned. A nonzero total chirality directly corresponds to the formation of unpaired Weyl points. Furthermore, by implementing a synthetic magnetic field via laser‑assisted tunneling in this setup, we demonstrate that the chirality imbalance drives a quantized charge current in the weak‑field regime, providing a direct signature of the CME. We verify that the adiabatic condition of the driving protocol, as well as the proposed experimental preparation and detection techniques, are within reach of current ultracold‑atom experiments. This work establishes a realistic and controllable platform for exploring chiral‑anomaly physics and nonequilibrium topological phenomena linked to Weyl fermions.
\end{abstract}

\maketitle

\section{Introduction}

Weyl fermions have played an important role in bridging high-energy and condensed-matter physics~\cite{Weyl1929,Jia2016,Armitage2018,Hasan2021}.
Although they have not been observed as elementary particles in nature, Weyl fermions can emerge as low‑energy quasiparticles in Weyl semimetals---a phase that has been realized in solid‑state systems~\cite{Lv2015a,Lv2015b,Xu2015,Lv2021} and simulated in photonic crystals~\cite{Lu2015,Noh2017} and ultracold-atom platforms~\cite{Wang2021}.
Remarkably, key properties of Weyl fermions, such as the chiral anomaly, are preserved in these condensed-matter analogs~\cite{Adler1969,Bell1969,Burkov2015}.
Research on Weyl semimetals has extensively focused on their topologically protected Fermi-arc surface states~\cite{Wan2011,HuangS2015,XuSY2016,Yang2017} and on anomalous electromagnetic responses rooted in the chiral anomaly, most notably the chiral magnetic effect (CME)~\cite{Fukushima2008,Zyuzin2012a,Goswami2013,Vazifeh2013,Kharzeev2014,Li2016}.

In conventional lattice systems, however, the Nielsen–Ninomiya theorem forces Weyl fermions to appear in pairs with opposite chiralities~\cite{Nielsen1981a,Nielsen1981b}, thereby forbidding a net electric current along an external magnetic field---namely, the CME. 
Recent theoretical advances show that this no‑go constraint can be circumvented in nonequilibrium settings~\cite{Budich2017,Sun2018,Higashikawa2019,Lee2019,Bessho2021,Kawabata2021,Zhu2023}.
In particular, three‑dimensional (3D) systems under adiabatic periodic driving can host unpaired Weyl points in their low‑energy quasienergy spectrum~\cite{Sun2018,Higashikawa2019}.
Such unpaired Weyl points are topologically protected by a nonzero winding number of the Floquet operator over the 3D Brillouin zone (BZ)~\cite{Kitagawa2010,Nathan2015}.
Despite these insights, existing proposals often require fine tuning of multiple parameters and involve complicated driving sequences~\cite{Higashikawa2019}, which can lead to considerable experimental overhead.
A practical, realistic protocol within a concrete quantum platform, accompanied by a detailed analysis of the resulting chiral magnetic response, therefore remains an open challenge.

In this work, we address the above challenge using ultracold atomic systems---an ideal platform for simulating topological states of matter due to their exceptional controllability and long coherence times~\cite{Goldman2016,ZhangDW2018,Cooper2019,Schafer2020}.
Specifically, optical Raman lattice (ORL) schemes~\cite{Liu2013,Liu2014,Wang2018,Lu2020}, in which a single set of laser fields simultaneously generates both an optical lattice potential and a spatially periodic Raman-coupling potential, offer a powerful route to realize diverse topological phases requiring spin‑orbit coupling (SOC). These include one-dimensional (1D) and two-dimensional (2D) topological insulators~\cite{Song2018b,Wu2016,SunW2018,Liang2023,Zhang2023} as well as 3D topological semimetals~\cite{Wang2021,Song2019}.
It has been proposed that the CME could be observed in an ORL scheme by engineering an energy offset between Weyl fermions of opposite chirality~\cite{Zheng2019,Lu2020}.

Here, we propose a realistic scheme to realize unpaired Weyl points with four-level ultracold atoms in a 3D periodically driven ORL. 
By engineering distinct relative symmetries between the lattice potential and four Raman potentials, this configuration generates an effective 3D SOC and realizes a tunable 3D topological insulator. Through synchronous, periodic, and adiabatic modulation of one Raman potential and the Zeeman term, 
we show that the Floquet quasienergy spectrum of the low-energy sector hosts eight Weyl points at high-symmetry momenta of the 3D BZ. 
Their total chirality can be precisely tuned, with a nonzero net chirality corresponding directly to the emergence of unpaired Weyl points.
The adiabaticity of the driving protocol is further verified by examining the quantization of spin pumping along momentum lines that pass through these Weyl points.
Finally, by introducing a synthetic magnetic field via laser-assisted tunneling, we demonstrate that the chirality imbalance drives the CME. 
This effect manifests as a quantized charge current in the weak-field regime and can be detected experimentally by tracking the center-of-mass displacement of the atomic cloud.
All elements of the proposed scheme are within reach of current ultracold-atom techniques, ensuring its experimental feasibility.

The remainder of this paper is organized as follows. 
In Sec.~\ref{Model}, we present the 3D ORL scheme, including the laser configuration, optical transitions, periodic driving protocol, and the resulting model Hamiltonian. 
In Sec.~\ref{Weyl Fermions}, we demonstrate the emergence of unpaired Weyl points in the low-energy quasienergy spectrum, determine their chiralities, and examine the condition for adiabatic driving. 
In Sec.~\ref{CME}, we show how a synthetic magnetic field can be generated within the same ORL setup and investigate the resulting CME. 
We conclude in Sec.~\ref{discussion} with a summary of the results and a discussion of experimental prospects and feasibility. 
Additional technical details and derivations are provided in the appendices.

\section{Optical Raman lattice scheme}\label{Model}

\begin{figure*}
	\includegraphics[width=0.75\textwidth]{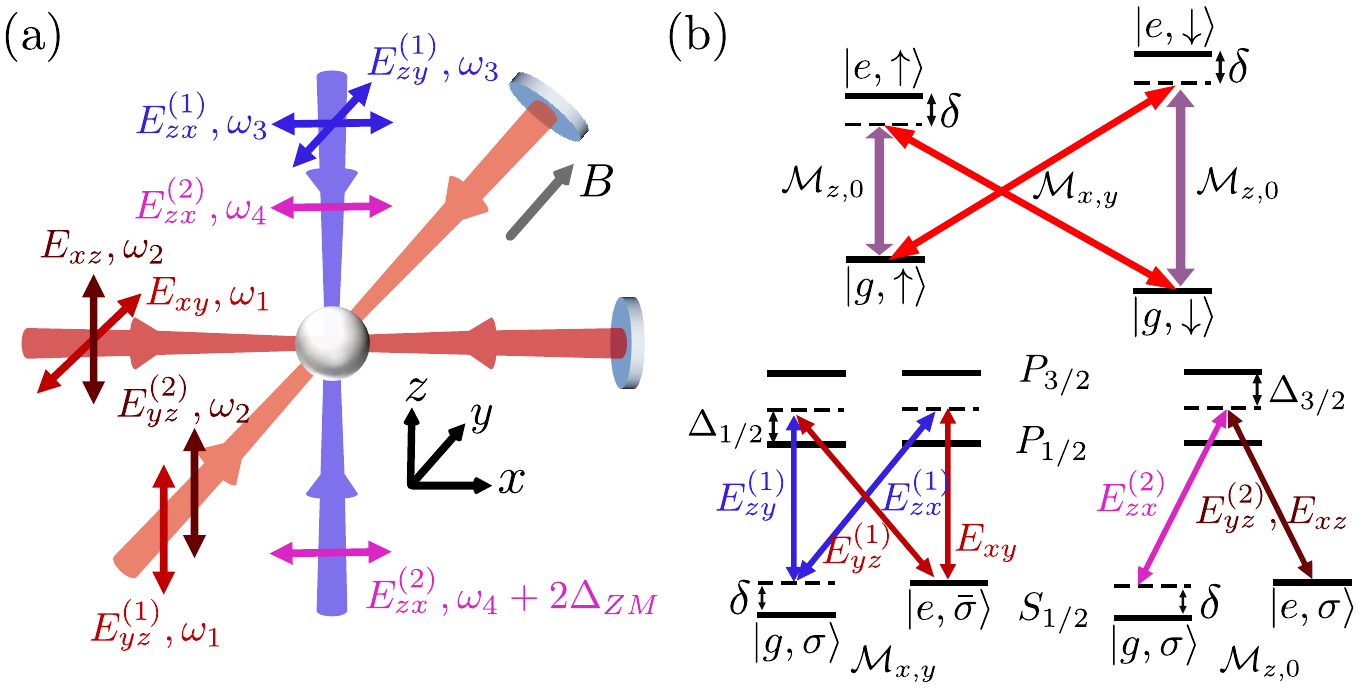}
	\caption{Schematic of the ORL scheme.
	(a) Experimental setup. A bias magnetic field is applied along the $y$ direction. In the $x$-$y$ plane, four incident laser beams are retroreflected by mirrors to form standing waves. Two beams at frequency $\omega_1$ are incident along $+x$ ($y$‑polarized, amplitude $E_{xy}$) and $+y$ ($z$‑polarized, amplitude $E_{yz}^{(1)}$), respectively, while the other two at frequency $\omega_2$ are both $z$‑polarized and incident along $+x$ (amplitude $E_{xz}$) and $+y$ (amplitude $E_{yz}^{(2)}$), respectively. Along the $z$ direction, three standing waves are created by counter‑propagating beam pairs: one at frequency $\omega_3$ containing both $x$ and $y$ polarization components (amplitudes $E_{zx}^{(1)}$ and $E_{zy}^{(1)}$), and two $x$‑polarized waves of equal amplitude $E_{zx}^{(2)}$ at frequencies $\omega_4$ and $\omega_4+2\Delta_{\text{ZM}}$, respectively.
	(b) Optical transitions in a four‑level system generating Raman potentials for alkali atoms.
	Top: States belonging to different orbital manifolds ($g$ and $e$) are coupled via two‑photon Raman transitions.
	Bottom: The spin-flip transitions $\ket{g,\sigma}\leftrightarrow\ket{e,\bar{\sigma}}$ are realized through double‑$\Lambda$ coupling configurations, which generate the potentials $\mathcal{M}_{x,y}$, whereas the spin-conserved transitions $\ket{g,\sigma}\leftrightarrow\ket{e,\sigma}$ are realized via single‑$\Lambda$ configurations, giving rise to the potentials $\mathcal{M}_{z,0}$. For alkali atoms, these potentials receive contributions from both the $D_1$ ($S_{1/2}\to P_{1/2}$) and $D_2$ ($S_{1/2}\to P_{3/2}$) lines. Here, $\sigma=\{\uparrow,\downarrow\}$ labels the spin state and $\bar{\sigma}$ denotes the opposite spin.
	}
	\label{fig1}
\end{figure*}

We aim to realize unpaired Weyl points for ultracold atoms in a 3D periodically driven ORL, as described by the Hamiltonian ($\hbar=1$)
\begin{align}
	\mathcal{H}(t) &= \left[\frac{ \mathbf{k}^2}{2m} + \mathcal{V}_{\text{latt}}({\bf{r}})\right] \otimes \id + \sum_{i=x,y,z} \mathcal{M}_i({\bf{r}}) \tau_x\otimes\sigma_i \nonumber\\
	&+ \mathcal{M}_{0}({\bf{r}},t) \tau_x\otimes\sigma_0 + m_z(t) \tau_z\otimes\sigma_0,
	\label{eq:HamLat}
\end{align}
which satisfies $\mathcal{H}(t)=\mathcal{H}(t+T)$. 
Here, $m$ is the atomic mass, $\mathbf{k}$ is the momentum, and $T$ denotes the driving period.
The Pauli matrices $\sigma_{x,y,z}$ and $\tau_{x,y,z}$ act on the spin and orbital (or pseudospin) subspaces, respectively. 
The identity matrices in these two subspaces are denoted by $\sigma_{0}$ and $\tau_{0}$, respectively, and the full identity operator in the combined spin-orbital space is given by $\id=\tau_0\otimes\sigma_0$. 
The state-independent lattice potential $\mathcal{V}_{\mathrm{latt}}(\mathbf{r})$ forms the underlying 3D lattice. 
Static, spatially periodic Raman potentials $\mathcal{M}_i(\mathbf{r})$ ($i=x,y,z$) are introduced to generate 3D SOC. 
Periodic driving is realized via synchronous modulation of the Raman potential $\mathcal{M}_{0}(\mathbf{r},t)$ and the Zeeman term $m_z(t)$.
In the tight-binding limit, $\mathcal{V}_{\mathrm{latt}}(\mathbf{r})$ and $\mathcal{M}_i(\mathbf{r})$ give rise to (pseudo)spin-conserved and (pseudo)spin-flipped nearest-neighbor hoppings, respectively, while $\mathcal{M}_{0}(\mathbf{r},t)$ drives on-site (pseudo)spin flips.
The explicit forms of $\mathcal{V}_{\mathrm{latt}}(\mathbf{r})$, $\mathcal{M}_i(\mathbf{r})$, $\mathcal{M}_{0}(\mathbf{r},t)$, and $m_z(t)$ will be specified below.

To implement the scheme, we employ four long-lived internal states of an alkali atom, denoted by $\ket{g,\sigma}$ and $\ket{e,\sigma}$ with $\sigma=\{\uparrow,\downarrow\}$. 
Here, $g$ and $e$ label two hyperfine manifolds of the electronic ground state, corresponding to the orbital (pseudospin) degrees of freedom, while $\uparrow$ and $\downarrow$ denote two Zeeman sublevels that together constitute an effective spin. 
As a concrete example, we consider $^{40}\mathrm{K}$ and select
$\ket{g,\uparrow} \equiv \ket{F=\tfrac{9}{2},\,m_F=\tfrac{1}{2}}$, $\ket{g,\downarrow} \equiv \ket{\tfrac{9}{2},-\tfrac{1}{2}}$, $\ket{e,\uparrow}\equiv \ket{\tfrac{7}{2},\tfrac{1}{2}}$ and $\ket{e,\downarrow}\equiv \ket{\tfrac{7}{2},-\tfrac{1}{2}}$. 
With appropriate choices of Zeeman sublevels, this scheme can be directly extended to a broad range of alkali atomic species.

We first outline the laser configuration; further technical details are provided in Appendix~\ref{App:Ham}. 
As shown in Fig.~\ref{fig1}(a), two beams at frequency $\omega_1$, with amplitudes $E_{xy}$ and $E^{(1)}_{yz}$, 
are incident along the $+x$ direction with $y$ polarization and the $+y$ direction with $z$ polarization, respectively. 
Both are retroreflected by mirrors to form standing waves. Setting $E^{(1)}_{yz}=E_{xy}$, the corresponding optical field reads
\begin{align}
		\mathbf{E}_1
		=&E_{xy}\left[\hat{\bf{y}}\sin(k_0 x)+\hat{\bf{z}}\sin(k_0 y)\right].
\end{align}
Here, $E_{\mu\nu}$ ($\mu,\nu\in\{x,y,z\}$) denotes the complex amplitude of the optical field propagating along the $\mu$ direction with $\nu$ polarization; superscripts distinguish different laser frequencies. 
A second pair of beams at frequency $\omega_{2}$, with amplitudes $E_{xz}$ and $E^{(2)}_{yz}$, is likewise retroreflected to create standing waves. Both are $z$-polarized, carry a relative phase of $\pi/2$, and are incident along the $+x$ and $+y$ directions, respectively. With $E^{(2)}_{yz}=E_{xz}$, this yields
\begin{align}
		\mathbf{E}_2=\hat{\bf{z}}E_{xz}\big[\cos(k_0 x)+\cos(k_0 y)\big].
\end{align}
The total optical field in the $x$-$y$ plane is $\mathbf{E}_{\textrm{$x$-$y$}}=\mathbf{E}_1+\mathbf{E}_2$.
The frequency difference $|\omega_1-\omega_2|$ is on the order of MHz, so these beams share approximately the same wave number $k_0$.
Along the $z$ direction, a standing wave at frequency $\omega_3$ is created by two counter-propagating beams. It contains both $x$ and $y$ polarization components with amplitudes $E^{(1)}_{zx}$ and  $E^{(1)}_{zy}$. Assuming $E^{(1)}_{zy}=E^{(1)}_{zx}$, we obtain
\begin{equation} \label{eq:Ez1}
	\mathbf{E}_3
	=\left(\hat{\bf{x}}+\hat{\bf{y}}\right)E^{(1)}_{zx}\cos(k_0 z).
\end{equation}
Additionally, two $x$-polarized standing waves of equal amplitude $E^{(2)}_{zx}$ are applied along $z$. 
Their frequencies are $\omega_4$ and $\omega_4 + 2\Delta_{\text{ZM}}$, respectively, with phases shifted by $\pm\eta(t)$. 
Here, $\Delta_{\text{ZM}}$ denotes the Zeeman splitting. Their combined field reads
\begin{align}\label{eq:Ez2}
	\mathbf{E}_4 = \hat{\bf{x}}E^{(2)}_{zx}\big\{\sin[k_0 z+\eta(t)]-\ui\sin[k_0 z-\eta(t)]\big\}.
\end{align}
The total optical field along $z$ is then $\mathbf{E}_z = \mathbf{E}_3 + \mathbf{E}_4$.
In the following, we assume $\eta(t)\ll 1$ and retain only first‑order terms in $\eta$.

The fields $\mathbf{E}_{\textrm{$x$-$y$}}$ and $\mathbf{E}_z$ introduced above simultaneously 
generate both a state-independent lattice potential and four Raman-coupling potentials (see Appendix~\ref{App:Ham} for details).
The lattice configuration in the $x$-$y$ plane is controlled by the two fields ${\bf E}_1$ and ${\bf E}_2$. The field ${\bf E}_1$ alone generates a regular square lattice. When ${\bf E}_2$ is turned on and adjusted so that $|E_{xz}|= |E_{xy}|$, the lattice transforms into a checkerboard pattern of depth $V_0\propto |E_{xz}|^2$, with neighboring sites oriented along the diagonal directions $x' = (x-y)/\sqrt{2}$ and $y'=(x+y)/\sqrt{2}$. This deformed lattice is essential for realizing the 3D SOC~\cite{Wang2021}. 
Along the $z$ direction, the standing waves $\mathbf{E}_3$ and $\mathbf{E}_4$ create a simple 1D lattice of depth $V_z\propto|E^{(1)}_{zx}|^2-|E^{(2)}_{zx}|^2$.
The total spin-independent optical lattice is therefore
\begin{equation}\label{eq: Vlatt}
\mathcal{V}_{\text{latt}}({\bf r})= -V_{0}\left[\cos^2(kx')+\cos^2(ky')\right]-V_z \cos^2(k_0z),
\end{equation}
where $k= k_0/\sqrt{2}$. 
The specific expressions for the lattice depths $V_0$ and $V_z$ in the case of $^{40}\mathrm{K}$ atoms are provided in Appendix~\ref{App:Ham}.
In the overall red-detuned regime both depths satisfy $V_0,V_z>0$.

Furthermore, the Raman potentials are generated by the combined action of two double-$\Lambda$ and two single-$\Lambda$ couplings induced by these optical fields, as illustrated in Fig.~\ref{fig1}(b).
The laser frequencies are chosen such that the difference $|\omega_3-\omega_1|$ nearly compensates the energy separation between $\ket{g,\sigma}$ 
and $\ket{e,\bar{\sigma}}$ (where $\bar{\sigma}$ denotes the opposite spin state), thereby inducing spin-flip transitions with a small two-photon detuning $\delta$. 
In contrast, the frequency pairs $(\omega_2,\omega_4)$ and $(\omega_2,\omega_4+2\Delta_{\text{ZM}})$ selectively address spin-conserved transitions $\ket{g,\uparrow}\leftrightarrow\ket{e,\uparrow}$ and $\ket{g,\downarrow}\leftrightarrow\ket{e,\downarrow}$, respectively, with the same two-photon detuning $\delta$.
In this setting, the two fields $\mathbf{E}_1$ $\mathbf{E}_3$ simultaneously form two double-$\Lambda$ couplings, which generate the static potentials
\begin{align}\label{eq: M_xy}
\begin{split}
	&\mathcal{M}_x({\bf r})= M_1 \sin(k x')\cos(k y')\cos(k_0 z),\\
	&\mathcal{M}_y({\bf r})=  - M_1 \cos(k x')\sin(k y')\cos(k_0 z),
\end{split}
\end{align}
with $M_{1}\propto E_{xy}^{*}E^{(1)}_{zx}$.
Meanwhile, the fields  $\mathbf{E}_2$ and $\mathbf{E}_4$ form two single-$\Lambda$ couplings, yielding the static potential 
\begin{align}\label{eq: M_z}
	\mathcal{M}_z({\bf r})= M_2 \cos(k x')\cos(k y')\sin(k_0 z),
\end{align}
and the time‑dependent potential
\begin{equation}\label{eq: M_0}
	\mathcal{M}_{0}({\bf r}, t)= \eta(t) M_2 \cos(k x')\cos(k y')\cos(k_0 z),
\end{equation}
with $M_2 \propto E^*_{xz}E^{(2)}_{zx}$.
The specific expressions for the Raman potential strengths $M_1$ and $M_2$ are provided in Appendix~\ref{App:Ham}.
The shared two-photon detuning $\delta$ results in a single Zeeman term $m_z=\delta/2$.

The total Hamiltonian thus takes the form of Eq.~\eqref{eq:HamLat}. Experimentally, the Zeeman term $m_z(t)$ can be tuned via the two‑photon detuning 
$\delta(t)$ using acousto-optic modulators~\cite{Zhang2023}, 
while $\eta(t)$ is controlled synchronously through the phases of the beams forming the field $\mathbf{E}_4$ with electro-optic modulators~\cite{Yi2019}.
Due to the spatial profiles of the Raman potentials $\mathcal{M}_{i}({\bf{r}})$ ($i=x,y,z$) relative to the lattice potential $\mathcal{V}_{\text{latt}}(\mathbf{r})$ [see Eqs.~\eqref{eq: Vlatt}-\eqref{eq: M_z}], each $\mathcal{M}_{i}({\bf{r}})$ is antisymmetric about every lattice site along the $i$ direction and symmetric along the other two directions.
This symmetry ensures that on‑site (pseudo)spin flips vanish, while giving rise to (pseudo)spin-flipped hopping between neighboring sites along the $i$ direction---effectively realizing a 3D SOC~\cite{Wang2021}. 
In contrast, the potential $\mathcal{M}_{0}(\mathbf{r},t)$ is fully symmetric about each lattice site [cf. Eqs.~\eqref{eq: Vlatt} and \eqref{eq: M_0}], and therefore generates only on-site (pseudo)spin flips.
For notational simplicity, we drop the primes hereafter and identify $(x,y,z)\equiv(x',y',z)$. Unless stated otherwise, all coordinates refer to this $(x,y,z)$ frame.

In the tight-binding limit and and considering only the $s$-bands, the system is described by the Bloch Hamiltonian (Appendix~\ref{App:Ham})
\begin{equation}
	H(\mathbf{k},t) = \sum_{l=0}^4 h_l(\mathbf{k},t)\,\gamma_l ,
	\label{eq:Hamq}
\end{equation}
where the coefficients $h_l(\mathbf{k},t)$ are
\begin{equation}
	\begin{aligned}
		h_0(\mathbf{k},t) &= m_z(t) - 2t_0 \sum_{i=x,y} \cos k_i-2t_z \cos k_z, \\
		h_1(\mathbf{k}) &= -2t_{\rm so}\sin k_x, \quad h_2 (\mathbf{k})=2t_{\rm so}\sin k_y, \\
		h_3(\mathbf{k}) &= -2t_{\rm soz}\sin k_z,\quad  h_4(t) =\eta(t) t_{\text{on}}.
	\end{aligned}
\end{equation}
The $\gamma$ matrices are defined as $(\gamma_0,\gamma_1,\gamma_2,\gamma_3,\gamma_4) = (\tau_z\otimes\sigma_0,\tau_y\otimes\sigma_x,\tau_y\otimes\sigma_y,\tau_y\otimes\sigma_z,\tau_x\otimes\sigma_0)$, which satisfy the Clifford algebra $\{\gamma_i,\gamma_j\}=2 \delta_{ij}\id$ ($i,j=0,1,\dots,4$). 
Here, $t_{0}$ and $t_{z}$ denote the (pseudo)spin-conserved hopping strengths in the $x$-$y$ plane and along the $z$ direction, respectively; 
$t_{\rm so}$ and $t_{\rm soz}$ are the corresponding (pseudo)spin‑flipped hopping strengths, and $t_{\mathrm{on}}$ is the on-site (pseudo)spin‑flip amplitude.
The Hamiltonian $H(\mathbf{k},t)$ describes a time-dependent 3D topological insulator, whose instantaneous eigenvalues are $E_{1,2}(\mathbf{k},t)=\mp\sqrt{\sum_{l=0}^{4}h_l^2(\mathbf{k},t)}$.

\section{Unpaired Weyl points}\label{Weyl Fermions} 

\begin{figure}
	\includegraphics[width=0.48\textwidth]{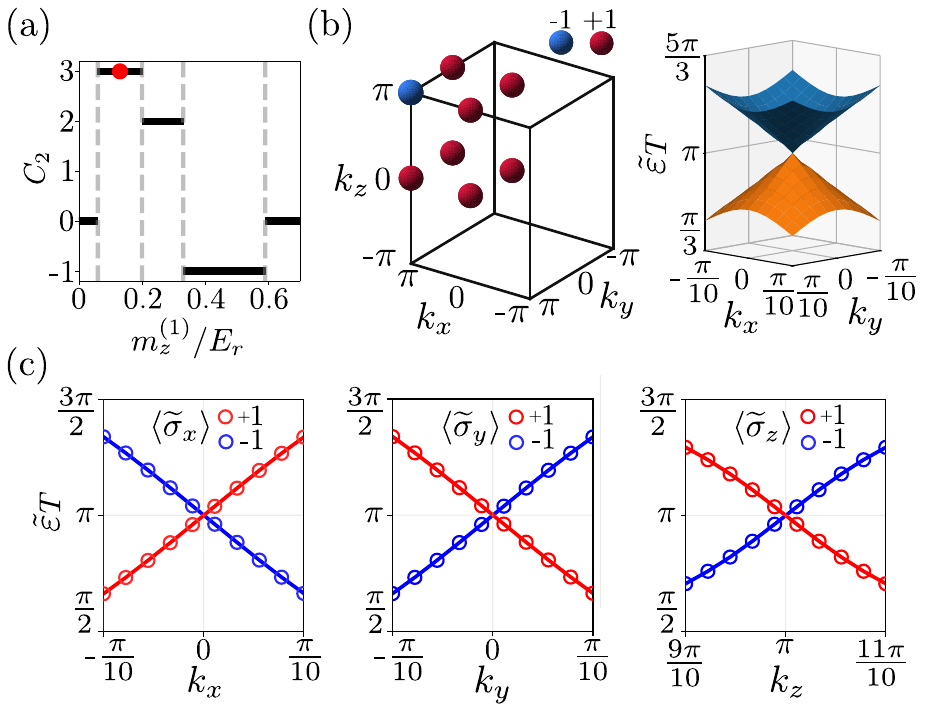}
	\caption{Unpaired Weyl points in the quasienergy spectrum of the low‑energy sector.
	(a) Phase diagram as a function of the mass parameter $m^{(1)}_z$. The red dot marks the value $m_z^{(1)}=0.13E_r$ used in (b)-(c).
	(b) Left: Locations and chiralities of the Weyl points in the 3D BZ.  
	Right: Quasienergy spectrum of the low‑energy Floquet operator $\tilde{U}_{\mathbf{k}}$ near the Weyl point located at $(0,0,\pi)$.
	(c) Spin texture of the low‑energy Floquet eigenstates around the Weyl point at $\mathbf{k}=(0,0,\pi)$, yielding $(\mu_x,\mu_y,\mu_z)=(1,-1,-1)$ and hence the chirality $\chi=\mu_x\mu_y\mu_z=+1$.
	All other parameters are fixed as $(V_0,V_z,M_1,M_2,m_z^{(2)})=(2,4,1,1,0.13)E_r$ and $\eta_0=0.15$.
	}\label{fig2}
\end{figure}

To realize unpaired Weyl points, a topologically nontrivial Floquet operator $U_{\mathbf{k}}={\cal T}\exp[-\ui\int_0^T H({\bf k},t) dt]$ (with ${\cal T}$ denoting time-ordering) is required~\cite{Sun2018,Higashikawa2019}, characterized by a nonzero 3D winding number
\begin{align}\label{eq: 3Dwinding}
\nu_3 = &\frac{1}{24\pi^2}\int d^{3}{\bf k}~\varepsilon^{\alpha \beta \gamma} \nonumber\\
&\times  \tr \left[(U_{\bf k}^{-1}\partial_{k_{\alpha}}U_{\bf k}) (U_{\bf k}^{-1}\partial_{k_{\beta}}U_{\bf k}) (U_{\bf k}^{-1}\partial_{k_{\gamma}}U_{\bf k})\right],
\end{align}
where $\varepsilon^{\alpha\beta\gamma}$ is the antisymmetric tensor with indices $\alpha,\beta,\gamma\!\in\!\{x,y,z\}$.
For a two-band Floquet operator, the total chirality of the Weyl points satisfies $\chi_{\text{tot}}=2\nu_3$~\cite{Sun2018}. 
To engineer topologically nontrivial $U_\mathbf{k}$, we initialize the system with the lower two bands fully occupied. 
Unless stated otherwise, this initial state---prepared in the static lattice before the periodic driving is turned on---is assumed throughout the rest of this work.

We then modulate the phase $\eta(t)$ and the Zeeman term $m_z(t)$ in the adiabatic limit as 
\begin{align}
\eta(t) = \eta_0 \sin(\omega t),\quad
m_z(t) = m^{(1)}_{z} - m^{(2)}_z\cos(\omega t),
\end{align}
where the driving frequency $\omega = 2\pi/T$ is taken to be much smaller than the gap of the instantaneous energy bands $E_{1,2}(\mathbf{k},t)$. 
The topology of $U_{\mathbf{k}}$ can be understood by interpreting the time $t$ as the fourth synthetic dimension.
In the adiabatic limit, the Bloch Hamiltonian maps onto a four-dimensional topological insulator model, whose topological invariant is the second Chern number $C_2$, equal to $\nu_3$~\cite{Qi2008,Kitagawa2010}. 
Adjusting the mass parameter $m_z^{(1)}$ allows the system to transition between various topological phases, with the resulting phase diagram displayed in Fig.~\ref{fig2}(a).

For the analysis in this section, we fix the static parameters to $V_0=2E_r$, $V_z=4 E_r$ and $M_1 = M_2 = 1 E_r$, where $E_r=k^2_0/(2m)$ is the recoil energy. 
For the periodic modulation, we choose $m_z^{(1)}= m_z^{(2)}=0.13 E_r$ and $\eta_0 = 0.15$, corresponding to the red point in Fig.~\ref{fig2}(a). 
Within the tight-binding description, these choices yield the hopping parameters
$(t_0,t_{z},t_{\rm so},t_{\rm soz},\eta_0t_{\text{on}}) \simeq (0.065,0.097,0.050,0.050,0.10)E_r$, as detailed in the Appendix~\ref{App:Ham}. For this parameter set, we have $C_2=\nu_3=3$, consistent with the net chirality of the unpaired Weyl points discussed below.

Now we demonstrate the emergence of unpaired Weyl points in the 3D BZ from the periodically driven Bloch Hamiltonian $H(\mathbf{k},t)$ and determine their chiralities.
The Bloch Hamiltonian $H(\mathbf{k},t)$ can be expressed in block-diagonal form within the eigenbasis of the spin sector. 
The two block Hamiltonians, acting on the orbital subspace, are given by $H_{\pm}(\mathbf{k},t) = h_0(\mathbf{k},t) \,\tau_z 
+ h_4(t)\,\tau_x \pm s(\mathbf{k}) \,\tau_y$, with $ s(\mathbf{k}) =  \sqrt{\sum_{l={1,2,3}} h^2_l(\mathbf{k})}$ (see Appendix~\ref{App:Bloch} for details).
Therefore, the two lowest bands of ${H}(\mathbf{k},t)$ evolve under $H_{+}(\mathbf{k},t)$ and $H_{-}(\mathbf{k},t)$, respectively. 
Importantly, $H_{+}$ and $H_{-}$ differ only by the sign of the $s(\mathbf{k})\,\tau_y$ term. 
As a result, the quasienergies of the two lowest Floquet bands become degenerate only at momenta where $s(\mathbf{k}) = 0$, i.e., at eight high-symmetry momenta $\mathbf{k}_j \in  \{ (k_x,k_y,k_z) | k_x,k_y,k_z \in \{0,\pi \} \} $ in the whole 3D BZ [see Fig.~\ref{fig2}(b)]. 
Projecting the Floquet operator $U_{\bf k}$ onto the low-energy subspace yields the time-evolution operator $\tilde{U}_{\bf k}=e^{-\ui \tilde{H}_F T}$ of this sector over
one driving period $T$, where $\tilde{H}_F $ is the effective low-energy two-band Hamiltonian.
Near each high-symmetry momentum $\mathbf{k}_j$, this low-energy effective Hamiltonian takes the Weyl form (see Appendix~\ref{App:Bloch})
\begin{align}
	\tilde{H}_F(\mathbf{k}) \propto \sum_{i=x,y,z} \mu_{ji}\big[k_i - (\mathbf{k}_{j})_i\big] \sigma_i, \label{eq: HF_low energy}
\end{align}
with the coefficients
\begin{align}
	\mu_{ji} =(-1)^{q_j/\pi}\ s_i\, \text{sgn}\big[\cos\, (\mathbf{k}_j)_i\big].  \label{eq: muji}
\end{align}
Here, $q_j = 0$ (or $\pi$) when the Weyl point lies in the quasienergy gap at $0$ (or $\pi/T$), and $s_x=-1$, $s_y=+1$, $s_z=-1$. 
The chirality of the $j$-th Weyl point is given by $\chi_j = \mu_{jx}\mu_{jy}\mu_{jz}$.

To verify this result, we also compute the quasienergy spectrum of the original ORL Hamiltonian $\mathcal{H}(t)$ near the high-symmetry momenta.
In the adiabatic limit, the low-energy Floquet operator $\tilde{U}_{\mathbf{k}}$ corresponds to a Wilson loop in parameter space~\cite{Wilczek1984,Bradlyn2022}
\begin{equation}\label{eq: low-energy Uk}
	\tilde{U}_{\mathbf{k}}
	= \mathcal{P}\exp\!\left[
	\ui \oint_{{\bm \Theta}(t)} \mathbf{a}_{\mathbf{k}}({\bm \Theta})\!\cdot d{\bm \Theta}
	\right],
\end{equation} 
where the closed path ${\bm \Theta}(t)$ represents the cyclic variation of the adiabatic parameters ${\bm \Theta}$ over $t\in[0,T]$, and $\mathcal{P}$ denotes path ordering. The non-Abelian Berry connection $\mathbf{a}_{\mathbf{k}}({\bm \Theta})$ has elements $\left[\mathbf{a}_{\mathbf{k}}({\bm \Theta})\right]_{mn}= \ui \langle \mathbf{k},{\bm \Theta},m| \nabla_{{\bm \Theta}}|\mathbf{k},{\bm \Theta},n\rangle$,
where $\ket{m}$ and $\ket{n}$ label the low-energy instantaneous eigenstates of $\mathcal{H}(t)$. 
Diagonalizing the operator $\tilde{U}_{\mathbf{k}}$ yields eigenvalues of the form $e^{-\ui\tilde{\varepsilon}_{n}(\mathbf{k})T}$ and the corresponding Floquet eigenstates $|\tilde{\psi}_n({\bf k})\rangle$ of the low-energy sector, with the quasienergies $\tilde{\varepsilon}_{n}(\mathbf{k})$ encoded in the phases. 
The coefficients $\mu_{ji}$ in Eq.~\eqref{eq: HF_low energy} can then be extracted---instead of using the result given in Eq.~\eqref{eq: muji}---from the spin polarizations $\langle\tilde{\sigma}_i\rangle$ of these Floquet eigenstates $|\tilde{\psi}_n({\bf k})\rangle$ evaluated near $\mathbf{k}_j$, as illustrated below.

The right panel of Fig.~\ref{fig2}(b) displays the quasienergy spectrum $\tilde{\varepsilon}_{n}(\mathbf{k})$ near the momentum $(k_x,k_y,k_z) =(0,0,\pi)$, clearly revealing the presence of Weyl points at this location.
To determine its chirality, we vary each momentum component $k_i$ ($i=x,y,z$) individually while fixing the other two at the Weyl point and compute the spin polarization $\langle\tilde{\sigma}_i\rangle$ [Fig.~\ref{fig2}(c)].
The quasienergy branches are color‑coded by $\langle\tilde{\sigma}_i\rangle$, with red (blue) corresponding to $\langle\tilde{\sigma}_i\rangle=+1$ ($-1$). 
The resulting spin texture yields $(\mu_x,\mu_y,\mu_z) = (1,-1,-1)$, giving the chirality $\chi = \mu_x \mu_y \mu_z = +1$.
The chiralities of all eight Weyl points are summarized in the left panel of Fig.~\ref{fig2}(b). The net chirality over the BZ is $\chi_{\mathrm{tot}} =\sum_j \chi_j =6$, which equals $2\nu_3$ as expected.

It is noteworthy that the quasienergy spectra and chiralities of the Weyl points obtained from the full ORL Hamiltonian $\mathcal{H}(t)$ agree with the analytical predictions based on the tight‑binding Bloch Hamiltonian $H(\mathbf{k},t)$. This agreement confirms that the proposed realization in a 3D periodically driven ORL faithfully implements the time‑dependent 3D topological insulator in the tight‑binding limit.

\begin{figure}
	\includegraphics[width=0.48\textwidth]{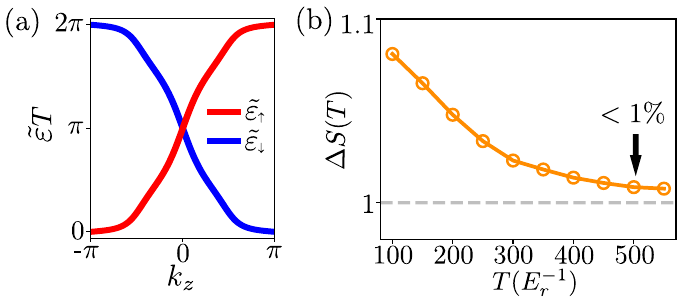}
	\caption{Spin pumping and adiabatic condition. 
	(a) Quasienergy spectrum of the low-energy sector along the momentum line $(k_x,k_y)=(0,\pi)$ exhibit opposite winding numbers in the two spin sectors, corresponding to pumped charges $Q_{\uparrow}=1$ (red) and $Q_{\downarrow}=-1$ (blue).
	(b) Pumped spin $\Delta S$ calculated from the full ORL Hamiltonian versus the driving period $T$. The dashed line marks the quantized value $\Delta S= 1$.
The deviation from quantization drops below $1\%$ for $T\geq500 E^{-1}_r$. All parameters are the same as in Fig.~\ref{fig2}. 
	}\label{fig3}
\end{figure}

We now examine the adiabatic condition of our driving protocol by analyzing the spin pumping process.
Specifically, we consider pumping along time-reversal-invariant momentum lines $(k_x,k_y) =(0,0)$, $(0,\pi)$, $(\pi,0)$ and $(\pi,\pi)$, which go through the Weyl points.
Along these lines, the spin-up and spin-down sectors decouple, allowing the spin pumping to be treated as two independent Thouless pumps. 
In the adiabatic limit, the charge pumped per driving cycle for spin $\sigma \in\{\uparrow,\downarrow\}$ is given by~\cite{Kitagawa2010,Privitera2018}
\begin{equation}\label{eq: Q_sigma}
	Q_{\sigma} = \frac{T}{2\pi} \int_{-\pi}^\pi dk_z\,
	\partial_{k_z} \tilde{\varepsilon}_\sigma(k_z) \in \mathbb{Z},
\end{equation}
where $\tilde{\varepsilon}_\sigma(k_z)$ denotes the quasienergies $\tilde{\varepsilon}_n(k_z)$ in the spin-$\sigma$ sector. 
The integral in Eq.~\eqref{eq: Q_sigma} measures the winding number of the quasienergies $\tilde{\varepsilon}_\sigma(k_z)$ as $k_z$ traverses the 1D BZ~\cite{Kitagawa2010}.
A non‑integer pumped charge therefore signals that a gap has opened at the Weyl points, preventing a full winding of the quasienergy bands.
This provides a direct diagnostic for nonadiabatic effects. 

To compute the net pumped spin per driving period $T$, we start from the eigenstates of the ORL Hamiltonian $\mathcal{H}(t=0)$, denoting those in the spin-$\sigma$ sector as $\ket{u_{\sigma}(k_z,0)}$. The total pumped spin is $\Delta S(T)= \int_{0}^{T} dt\, \langle J_{s}(t)\rangle$, where the spin current is defined as
	\begin{align}
		\langle J_{s}(t)\rangle =& \frac{1}{2\pi} \int_{-\pi}^{\pi} dk_z
		\left[\bra{u_{\uparrow}(k_z,t)} \,\partial_{k_z}\mathcal{H}(t)\, \ket{u_{\uparrow}(k_z,t)}\right.\nonumber\\
		&\left.-\bra{u_{\downarrow}(k_z,t)} \,\partial_{k_z}\mathcal{H}(t)\, \ket{u_{\downarrow}(k_z,t)}\right],
	\end{align}
with $\ket{u_{\sigma}(k_z,t)}={\cal T}e^{-\ui\int_0^t {\cal H}(t') dt'}\ket{u_{\sigma}(k_z,0)}$. 
We focus on spin pumping along the line $(k_x,k_y)=(0,\pi)$, where the minimal instantaneous gap (over $k_z$) is the smallest among all high-symmetry lines, thereby imposing the most stringent constraint on the driving period.
As shown in Fig.~\ref{fig3} (a), the quasienergies $\tilde{\varepsilon}_n(k_z)$ in spin-up (spin-down) sector for $(k_x,k_y)=(0,\pi)$ are calculated from $\tilde{U}_{\mathbf{k}}$ using Eq.~\eqref{eq: low-energy Uk} and are plotted as red (blue) curves, yielding the corresponding pumped charges $Q_{\uparrow} = +1$ and $Q_{\downarrow} = -1$. This results in a quantized spin pump per cycle in the adiabatic limit,
\begin{equation}
	\Delta S(T \to \infty) = \frac{Q_{\uparrow} - Q_{\downarrow}}{2} = 1.
\end{equation}
In Fig.~\ref{fig3} (b), we plot $\Delta S(T)$ as a function of the driving period $T$. The spin pumped per cycle deviates from the quantized value $\Delta S(T \to \infty)= 1$ by less than $1\%$ for $T\geq500 E^{-1}_r$; in this regime, the adiabatic condition can be taken as satisfied. 
Experimentally, the pumped charges at selected momenta $(k_x,k_y)$ can be extracted from the spin-resolved center-of-mass displacement along the $z$ direction by combining time-of-flight expansion only in the $x$-$y$ plane~\cite{Wang2013} with slice-selective imaging~\cite{Sunami2022}.

\section{Chiral magnetic effect}\label{CME}

\begin{figure}
	\includegraphics[width=0.48\textwidth]{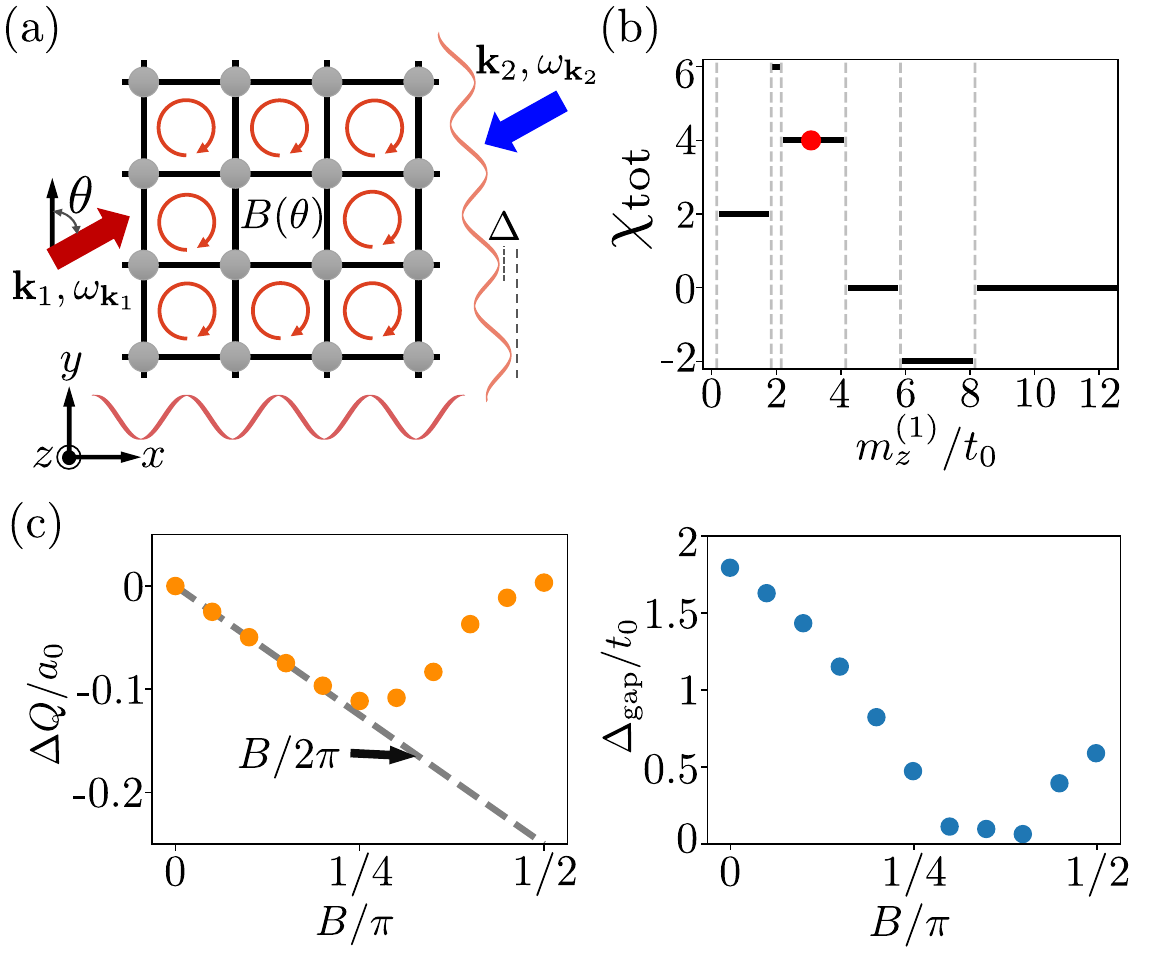}
	\caption{Quantized charge current as a manifestation of the CME.
	(a) Laser-assisted tunneling scheme. A linear tilt potential $\Delta$ along $y$ suppresses the bare hopping, which is resonantly restored by two running-wave beams 
with wavevectors $\mathbf{k}_1, \mathbf{k}_2$ and frequencies $\omega_1,\omega_2$.
	(b) Phase diagram as a function of the mass parameter $m^{(1)}_z$ for $\theta=0$. The red dot marks the value $m_z^{(1)}=0.2E_r$ used in (c).
	(c) Left: Normalized pumped charge $\Delta Q$ per driving cycle versus magnetic‑field strength $B$. Right: Instantaneous energy gap between the low‑ and high‑energy bands as a function of $B$.
	All other parameters are taken as $(t_0,t_{z},t_{\rm so},t_{\rm soz}, \eta_0t_{\text{on}}, m_z^{(2)}) = (0.065,0.097,0.050,0.050,0.10,0.13)E_r$, with the driving period $T=500 E_r^{-1}$}\label{fig4}
\end{figure}

In the presence of a magnetic field, the chiral anomaly of a Weyl fermion gives rise to a charge current parallel to the field, a response known as the CME~\cite{Fukushima2008}. In this section, we introduce a synthetic magnetic field and numerically demonstrate that the charge current can be observed within our periodically modulated ORL setup.

Synthetic magnetic fields acting on neutral atoms can be engineered through laser-assisted tunneling~\cite{Aidelsburger2013,Miyake2013,Aidelsburger2015,Kennedy2015}. As illustrated in Fig.~\ref{fig4}(a), a spin-independent linear potential of amplitude $\Delta$ between neighboring sites along the $y$ direction can be created using either an ac Stark shift gradient~\cite{Dimitrova2020} or accelerated optical lattice~\cite{Dahan1996,Wilkinson1996,Sias2007}. 
When $\Delta \gg t_0$, tunneling along $y$ is strongly suppressed but can be resonantly restored by applying a pair of far-detuned running-wave beams with wave vectors
$\mathbf{k}_1$ and $\mathbf{k}_2$ and a frequency difference $\omega_{\mathbf{k}_1} - \omega_{\mathbf{k}_2}= \Delta$. 
These beams generate a local optical potential $V(\mathbf{r}) = V_K \cos^2\!\big(\delta \mathbf{k} \cdot \mathbf{r}/2 + \Delta t/2\big)$ with $\delta \mathbf{k} = \mathbf{k}_1 - \mathbf{k}_2$. In the high-frequency limit $\Delta \gg t_0$, atomic hopping along the $y$ direction acquires a site-dependent Peierls phase factor $\phi_{m,n,l} = \delta \mathbf{k} \cdot \mathbf{R}$ (up to a constant phase shift; see Appendix~\ref{App:Ham_with_B}), where $\mathbf{R} =  a_0(\sqrt{2}m\hat{\mathbf x}+ \sqrt{2} n\hat{\mathbf y}+ l\hat{\mathbf z})$ denotes the position of the lattice site ${\bf j}\equiv(m,n,l)$, with $a_0 = \pi/k_0$. 
We set the two beams to counter-propagate within the $x$-$y$ plane, with wave vectors $\mathbf{k}_{1}=-\mathbf{k}_{2}$ and wavelength $\lambda$. 
Their common propagation direction makes an angle $\theta$ with respect to the $y$ axis [Fig.~\ref{fig4}(a)]. This configuration yields the site-dependent phase $\phi_{m,n,l} = 2\pi\beta (m \sin\theta + n\cos\theta)$ with $\beta = 2\sqrt{2}a_0/\lambda$, corresponding to a uniform magnetic field of magnitude $|B| = 2\pi\beta\sin\theta$ along the $-z$ direction. 
The resulting effective tight-binding Hamiltonian reads (see Appendix~\ref{App:Ham_with_B} for the derivation)
\begin{align}\label{eq:Heff_Azx}
	H_{\rm eff} =&
	\sum_{m,n,l} \Big\{\Big [
	C_{m+1,n,l}^\dagger\big(-t^x_0 \tau_z\sigma_0+ \ui s_x t_{\rm so}^{x}\tau_y\sigma_x\big)C_{m,n,l} \nonumber\\
	&+C_{m,n+1,l}^\dagger\, e^{-\ui \phi_{m,n,l}}\big(-t^y_0\tau_z\sigma_0+ \ui s_y t_{\rm so}^{y} \tau_y\sigma_y\big)C_{m,n,l} \nonumber\\
	& +C_{m,n,l+1}^\dagger\big(-t_z\tau_z\sigma_0+ \ui s_z t_{\rm soz}\tau_y\sigma_z\big)C_{m,n,l}+ \mathrm{H.c.}\Big] \nonumber\\
	&+C_{m,n,l}^\dagger \big[m_z(t)\,\tau_z\sigma_0 + \eta(t)\,t_{\rm on}\,\tau_x\sigma_0\big]C_{m,n,l}\Big\},
\end{align}
where $C_{\bf j}\equiv (c_{{\bf j}, g\uparrow}, c_{{\bf j}, g\downarrow},c_{{\bf j}, e\uparrow},c_{{\bf j}, e\downarrow})^{\rm T}$ 
denotes the vector of annihilation operators for all internal states at site $\mathbf{j}$, with $c_{{\bf j}, \alpha\sigma}$ ($c_{{\bf j}, \alpha\sigma}^\dagger$) annihilating (creating) an atom in the state $\ket{\alpha,\sigma}$ ($\alpha\in\{g, e\}$) at that site. 
Here, the photon-assisted tunneling amplitudes are given by (Appdenix~\ref{App:Ham_with_B})
\begin{align}
	\begin{split}
	t^{x,y}_0  &= t_0\,\mathcal{J}_{0,1}\!\left(\frac{V_K}{\Delta}\sin\kappa_{1,2}\right), \\
	t^{x,y}_{\rm so}   &= t_{\rm so}\,\mathcal{J}_{0,1}\!\left(\frac{V_K}{\Delta}\sin\kappa_{1,2}\right),
	\end{split}
\end{align}
with $\mathcal{J}_n$ denoting the $n$-th order Bessel function of the first kind, $\kappa_1(\theta)\equiv \pi\beta\sin\theta$, and $\kappa_2(\theta)\equiv \pi\beta\cos\theta$.
The strength of the synthetic magnetic field can be tuned continuously by adjusting the angle $\theta$.

Here, we set $|\mathbf{k}_1| = k_0$, which gives $\beta = \sqrt{2}$, and fix the ratio $V_K/\Delta=2$. All other parameters are the same as those in Sec.~\ref{Weyl Fermions}, except for the now tunable parameter $m_z^{(1)}$. This adjustment is necessary because laser-assisted tunneling modifies the effective hopping amplitudes within the $x$-$y$ plane, which in turn alters the system's phase diagram. As shown in Fig.~\ref{fig4}(b), we plot the phase diagram as a function of $m_z^{(1)}$ for $\theta=0$ ($B=0$). 
Accordingly, we take $m_z^{(1)}=0.2 E_r$ for the subsequent analysis of the CME, corresponding to a total chirality of the Weyl fermions of $\chi_{\text{tot}}=4$.

To compute the induced charge current, we adopt the Landau gauge $\mathbf{A} = (0,Bm,0)$, so that $k_y$ and $k_z$ remain good quantum numbers. 
The eigenstates of $H_{\text{eff}}$ at the initial time can be labeled $\ket{u_{b}(k_y,k_z,t=0)}$, where $b$ indexes the discrete Landau levels along $x$. 
When the lower bands are initially filled, the total charge pumped along $z$ over one driving period $T$ is $Q(T) = \int_0^T dt\, \langle J_z(t)\rangle$, 
where the charge current reads
\begin{align}
	 \langle J_z(t)\rangle=&\sum_{b} \frac{1}{4\pi^2}\int^{\pi}_{-\pi} \int^{\pi}_{-\pi} dk_y dk_z \nonumber\\
	&\times \bra{u_{b}(k_y,k_z,t)} \partial_{k_z} H_{\rm eff}(t)\ket{u_{b}(k_y,k_z,t)},
\end{align}
with $\ket{u_{b}(k_y,k_z,t)}={\cal T}e^{-\ui\int_0^t H_{\rm eff}(t') dt'}\ket{u_{b}(k_y,k_z,0)}$. 
As shown in the left panel of Fig.~\ref{fig4}(c), we compute the pumped charge per cycle, normalized by the number of occupied Landau-level states,
$\Delta Q=Q(T)/N_{\rm LL}$, as a function of the magnetic-field strength $B$ for $T=500E_r^{-1}$.
In the weak-field regime, the pumped charge follows a quantized relation proportional to the total chirality:
\begin{align}
\Delta Q =\frac{\chi_{\mathrm{tot}}B}{8\pi},  \label{eq: Delta Q}
\end{align}
which gives $\Delta Q =B/(2\pi)$ for $\chi_{\mathrm{tot}}=4$.

This quantization has a topological origin. In the adiabatic limit, 
\begin{align}
	Q(T\to\infty)=\sum_{b}\frac{T}{4\pi^2} \int^{\pi}_{-\pi} \int^{\pi}_{-\pi} dk_y dk_z
	\partial_{k_z} \tilde{\varepsilon}_{b}(k_y,k_z) 
\end{align}
where $\tilde{\varepsilon}_{b}(k_y,k_z)$ are the quasienergy bands of the low-energy sector under the synthetic magnetic field. 
Hence, the pumped charge $Q$ corresponds to the winding number of $\tilde{\varepsilon}_{b}(k_y,k_z)$ along $k_z$.
The only nonvanishing contribution to $Q$ comes from the paired zeroth Landau levels of the same chirality, located in the $0$ and $\pi$ quasienergy gaps, respectively.
Consequently, $Q(T\to\infty)= \Phi L_x L_y \chi_{\mathrm{tot}}/2$, where $\Phi L_x L_y$ counts the number of the zeroth-Landau-level states, with $\Phi={B}/2\pi$ being the flux per plaquette and $L_x$ ($L_y$) the number of sites along $x$ ($y$). 
Using the total number of occupied Landau-level states $N_{\rm LL}=2L_xL_y$, the normalized pumped charge $\Delta Q = Q(T)/N_{\rm LL}$ must satisfy the quantized relation in Eq.~\eqref{eq: Delta Q} in the adiabatic limit.

Beyond a critical field strength $B_c\approx\pi/4$, however, the quantization breaks down. 
This occurs because the magnetic field closes the gap between the low- and high-energy bands, thereby violating the adiabatic approximation.
As demonstrated in the right panel of Fig.~\ref{fig4} (c), the interband gap $\Delta_{\text{gap}}$ vanishes around $B_c$.

\section{Conclusion and discussion}\label{discussion}

In summary, we have proposed a Floquet-engineered 3D ORL scheme for realizing unpaired Weyl points via adiabatic periodic modulation of a 3D tight-binding topological insulator. The Weyl points emerge in the quasienergy spectrum of the low-energy sector, and their total chirality can be tuned to distinct nonzero values, corresponding to a controllable chirality imbalance. Furthermore, a synthetic magnetic field can be implemented in this setup using laser-assisted tunneling. We demonstrate that the resulting CME manifests as a time-dependent charge current parallel to the magnetic field, with the pumped charge per cycle being quantized to the total chirality in the weak-field regime.

The proposed scheme is experimentally feasible with current ultracold‑atom techniques. 
First, ORL schemes have already been implemented in both alkali~\cite{Wu2016,SunW2018,Wang2021,Zhang2023} and alkaline-earth~\cite{Song2018b,Song2019,Liang2023} atoms. Floquet engineering has also been applied to an ORL to realize tunable anomalous topological bands~\cite{Zhang2023}.
For $^{40}\mathrm{K}$ atoms, an ORL can be realized using laser beams at a wavelength of 768\,nm~\cite{Huang2016}; the corresponding recoil energy is $E_r=h\times8464$\,Hz.
Second, as shown in Fig.~\ref{fig3}(b), a driving period of $T=500 E_r^{-1}$ is sufficient to satisfy the adiabatic condition. 
This timescale is well below the typical lifetime of ultracold $^{40}\mathrm{K}$ atoms in a 2D ORL, which is on the order of $\tau\sim1300E_r^{-1}$ (about 150\,ms~\cite{Wang2018}), making adiabatic periodic modulation practicable~\cite{footnote}.
Third, laser-assisted tunneling is an established technique for realizing synthetic magnetic fields in ultracold gases~\cite{Aidelsburger2013,Miyake2013,Aidelsburger2015,Kennedy2015}. 
The field strength can be tuned continuously by varying the angle $\theta$, and the resulting charge current can be detected by monitoring the center‑of‑mass displacement of the atomic cloud~\cite{Lohse2016,Nakajima2016}.
Furthermore, a mean-field analysis for a representative inter-state scattering length in ultracold $^{40}\mathrm{K}$ atoms indicates that the on-site interaction lies in the weak-to-intermediate regime ($U\sim t_0$) and mainly renormalizes the driving parameters; this does not hinder the realization of unpaired Weyl points. However, interactions can enhance Floquet heating and thus reduce the coherence time, which should be carefully evaluated in future experiments.

Before closing, we briefly discuss several open questions that the proposed setup is uniquely positioned to address. In the absence of periodic driving, the system realizes a 3D chiral topological insulator whose four internal states naturally form the sublattice basis for chiral symmetry. This opens the door to studying interaction-driven chiral-symmetry-breaking transitions that could occur without closing the single-particle gap---a type of symmetry-breaking-driven topological transition that has remained experimentally elusive. 
With a sharp boundary generated by projecting a repulsive potential step using a programmable digital micromirror device~\cite{Braun2024}, one could directly resolve the boundary spin textures and benchmark their robustness against controlled symmetry-breaking perturbations. Under periodic driving, the emergence of unpaired Weyl points and the CME provides a controlled Floquet setting for investigating chiral anomaly transport~\cite{Kharzeev2014} and its dependence on driving parameters and interactions. Furthermore, the well-defined pseudospin structure enables the realization of non-Hermitian chiral topological matter by introducing state-selective atomic loss~\cite{Zhao2025}, granting access to phenomena such as chiral-symmetry-protected exceptional surfaces, the interplay between Weyl points and the non-Hermitian skin effect, and the breakdown of bulk-boundary correspondence in three dimensions~\cite{Bergholtz2021}. Together, these perspectives highlight the potential of our scheme as a versatile simulator for topological physics beyond the reach of current experiments.

\begin{acknowledgments}
This work was supported by the National Natural Science Foundation of China (Grant No. 12574294, No. 12374248, and No. 12204187), and the Quantum Science and Technology-National Science and Technology Major Project (Grant No. 2021ZD0302000). J. Z. acknowledges support from the Fundamental Research Funds for the Central Universities 
(Grant No. WK9990000122).
\end{acknowledgments}

\section*{Data Availability}

The data that support the findings of this article are openly available~\cite{Data}.

\appendix
\section{Details of experimental realization for $^{40}$K atoms}\label{App:Ham}

In this Appendix, we take $^{40}$K atoms as a concrete example and detail the experimental realization of the ORL scheme introduced in Sec.~\ref{Model}.

\subsection{Optical fields}

The laser configuration is illustrated in Fig.~\ref{fig1}(a).
In the $x$-$y$ plane, four incident laser beams are retroreflected by mirrors to form standing waves.
Two beams at frequency $\omega_1$ are incident along the $+x$ direction with $y$ polarization and the $+y$ direction with $z$ polarization, respectively.
Their combined field is
\begin{align}
\mathbf{E}_1=\,&\hat{\bf{y}} |E_{xy}| e^{\ui(\alpha'+\alpha'_{L}/2)} \cos(k_0 x-\alpha'_L/2)\nonumber\\
		&+\hat{\bf{z}} |E^{(1)}_{yz}| e^{\ui(\beta'+\beta'_{L}/2)} \cos(k_0 y-\beta'_L/2).
\end{align}
The other two beams, both at frequency $\omega_2$  and  $z$-polarized, are incident along $+x$ and $+y$, respectively, giving
\begin{align}\label{eq: Exy_App}
	\mathbf{E}_2=& \hat{\bf{z}}|E_{xz}| e^{\ui(\alpha+\alpha_L/2)}\cos(k_0 x-\alpha_L/2)\nonumber\\
		&+\hat{\bf{z}}|E^{(2)}_{yz}|e^{\ui(\beta+\beta_L/2)}\cos(k_0 y-\beta_L/2).
\end{align}
Here $E_{\mu\nu}$ ($\mu,\nu\in\{x,y,z\}$) denotes the complex amplitude, $\alpha$, $\alpha'$, $\beta$, $\beta'$ are the initial phases,
and $\alpha_L$ ($\beta_L$) is the additional phase accumulated by the component $E_{xz}$ ($E^{(2)}_{yz}$) after traveling from the atoms to the mirror and back in the $x$ ($y$) direction.  The phases $\alpha'_L$, $\beta'_L$ have the same meaning for the components $E_{xy}$ and $E^{(1)}_{yz}$, respectively.
Because of the frequency difference $|\omega_1-\omega_2|$, the distances between the atoms and the mirrors can be adjusted to tune these accumulated phases~\cite{Wu2016}. 
They can be set to satisfy (with $n\in\mathbb{Z}$):
\begin{align}
\alpha'_L-\alpha_L=(2n+1)\pi,\quad\beta'_L-\beta_L=(2n+1)\pi,
\end{align}
which yields the total optical field in the $x$-$y$ plane:
\begin{align}
	\begin{split}
	&\mathbf{E}_{\textrm{$x$-$y$}}=\mathbf{E}_1+\mathbf{E}_2, \\
	&\mathbf{E}_1=\, \hat{\bf{y}} |E_{xy}| e^{\ui(\alpha'+\alpha_{L}/2+\pi/2)} \sin(k_0 x-\alpha_L/2)\\
		&\quad\quad\,+\hat{\bf{z}} |E^{(1)}_{yz}| e^{\ui(\beta'+\beta_{L}/2+\pi/2)} \sin(k_0 y-\beta_L/2), \\
	&\mathbf{E}_2=\hat{\bf{z}}|E_{xz}| e^{\ui(\alpha+\alpha_L/2)}\cos(k_0 x-\alpha_L/2)\\
		&\quad\quad\,+\hat{\bf{z}}|E^{(2)}_{yz}|e^{\ui(\beta+\beta_L/2)}\cos(k_0 y-\beta_L/2).
	\end{split}
\end{align}

Along the $z$ direction, three standing waves are created by counter‑propagating beam pairs. 
The first, at frequency $\omega_3$, contains both $x$ and $y$ polarization components:
\begin{align}
	\mathbf{E}_3=\,&\hat{\bf{x}} |E^{(1)}_{zx}| e^{\ui  (\vartheta+\vartheta')/2 }\cos[k_0 z +(\vartheta-\vartheta')/2]\nonumber\\
	&+ \hat{\bf{y}} |E^{(1)}_{zy}| e^{\ui  (\vartheta+\vartheta')/2 }\cos[k_0 z +(\vartheta-\vartheta')/2].
\end{align}
The other two are $x$‑polarized standing waves of equal amplitude, with frequencies $\omega_4$ and $\omega_4+2\Delta_{\text{ZM}}$, respectively:
\begin{align}
	\mathbf{E}_4
	=&\hat{\bf{x}} |E^{(2)}_{zx}|\big\{ e^{\ui(\varphi_1 +\varphi_1')/2}
	\cos[k_0 z +(\varphi_1-\varphi_1')/2]\nonumber\\
	&+e^{\ui(\varphi_2+ \varphi_2')/2}\cos[k_0 z+(\varphi_2-\varphi_2')/2]\big\}.
\end{align}
Here, $\vartheta$ and $\vartheta'$ are the initial phases of the forward and backward beams that form $\mathbf{E}_3$; 
$\varphi_{1,2}$ and $\varphi'_{1,2}$ play analogous roles for the two components of $\mathbf{E}_4$.
Note that the first component of $\mathbf{E}_4$ (associated with phases $\varphi_{1}$, $\varphi'_{1}$) operates at frequency $\omega_4$, 
whereas the second component (associated with phases $\varphi_{2}$, $\varphi'_{2}$) is at frequency $\omega_4+2\Delta_{\text{ZM}}$.
Experimentally, the relative phases of $\mathbf{E}_3$ and $\mathbf{E}_4$ can be tuned dynamically via electro‑optic modulators (EOMs).
In the following, we impose the phase relations
\begin{align}
	\begin{split}
	&\varphi_1-\vartheta=\eta(t)-\pi/2,\quad \varphi'_1-\vartheta'=-\eta(t)+\pi/2, \\
	&\varphi_2-\vartheta=-\eta(t)-\pi,\quad \varphi'_2-\vartheta'=\eta(t),
	\end{split}
\end{align}
which yield the total field along the $z$ direction:
\begin{align}
	\begin{split}
	&\mathbf{E}_{z} = \mathbf{E}_3+\mathbf{E}_4, \\
 	&\mathbf{E}_3=\big(\hat{\bf{x}} |E^{(1)}_{zx}|+\hat{\bf{y}} |E^{(1)}_{zy}|\big)e^{\ui  \phi'_z }\cos(k_0 z +\phi_z), \\
 	&\mathbf{E}_4=\hat{\bf{x}} |E^{(2)}_{zx}|e^{\ui \phi'_z}\big\{\sin[k_0 z +\phi_z+\eta(t)] \\
		&\quad\quad\, -\ui\sin[k_0 z+\phi_z-\eta(t)]\big\},
	\end{split}
\end{align}
where we have defined $\phi_z\equiv(\vartheta-\vartheta')/2$ and $\phi'_z\equiv(\vartheta+\vartheta')/2$.

\subsection{Lattice and Raman potentials}

We now derive the lattice and Raman potentials generated by the optical fields described above. Several preliminaries must first be established.
First, for $^{40}\mathrm{K}$ atoms, the laser wavelength is typically set to $\lambda_0=768$\,nm, giving the wave number $k_0=2\pi/\lambda$.
The laser frequencies lie between the  $D_1$ ($4^{2}S_{1/2}\to4^{2}P_{1/2}$) and $D_2$ ($4{^{2}S}_{1/2}\to4{^{2}P}_{3/2}$) lines; consequently, both transitions contribute to the lattice and Raman potentials, as illustrated in Fig.~\ref{fig1}(b). We denote the detunings from these resonances by $\Delta_{1/2}$ and $\Delta_{3/2}$, respectively.
Second, with a bias magnetic field applied along the $y$ direction [Fig.~\ref{fig1}(a)], the spherical basis vectors are
$\hat{\bf e}_{0}=\hat{\mathbf{y}}$ and $\hat{\bf e}_{\pm}=\mp(\hat{\mathbf{z}}\pm \ui\hat{\mathbf{x}})/\sqrt{2}$.
Finally, we assume equal field amplitudes $|E_{xy}| = |E^{(1)}_{yz}| = |E_{xz}| = |E^{(2)}_{yz}|$ and $|E^{(1)}_{zx}|=|E^{(1)}_{zy}|$.

To obtain the lattice potential, contributions from all beams and from all relevant $\pi$- and $\sigma^{\pm}$-transitions must be summed. 
The lattice potential for a given internal state $\lvert\alpha,\sigma\rangle$ (with $\alpha \in \{g,e\}$ and $\sigma \in \{\uparrow,\downarrow\}$) is derived as
\begin{align}\label{eq:Potentials}
	{\cal V}_{\alpha\sigma}=\sum_{J=1/2,3/2} \sum_{F}\sum_{\zeta=1,2,3,4}\sum_{\nu=0,\pm}\frac{\big|\Omega^{(J)}_{\alpha\sigma,F;\zeta,\nu}\big|^{2}}{\Delta_J},
\end{align}
where the Rabi frequencies are given by
\begin{align}
	\begin{split}
	\Omega^{(J)}_{\alpha\sigma,F;\zeta,0}&=\langle\alpha,\sigma |er| F, m_{F\sigma},J \rangle \hat{\bf e}_{0}\,\cdot {\bf E}_\zeta,\\
	\Omega^{(J)}_{\alpha\sigma,F;\zeta,\pm}&=\langle\alpha,\sigma |er| F, m_{F\sigma}\pm1,J \rangle \hat{\bf e}_{\pm}\,\cdot {\bf E}_\zeta ,
	\end{split}
\end{align}
with $m_{F\sigma}$ denoting the magnetic quantum number of the state $\lvert\alpha,\sigma\rangle$.
Using the dipole matrix elements of $^{40}\mathrm{K}$~\cite{Tiecke2011}, we obtain a state-independent lattice potential
\begin{align} \label{eq: V_alpha sigma_00}
{\cal V}_{\alpha\sigma} =&-V_0 \cos(\delta\phi)\cos(k_0x-\alpha_L/2)\cos(k_0y-\alpha_L/2)\nonumber\\
 &-V_z \cos^2(k_0z+\phi_z)+{\cal O}(\eta^2),
\end{align}
where $\delta \phi = \alpha+\alpha_L/2 - \beta - \beta_L/2$, and
\begin{align}
	\begin{split}
	V_0 &=-\frac{2}{3}\left(\frac{t^2_{3/2}}{\Delta_{3/2}}+\frac{t^2_{1/2}}{\Delta_{1/2}}\right) |E_{xz}|^2, \\
	V_z &= -\frac{2}{3}\left(\frac{t^2_{3/2}}{\Delta_{3/2}}+\frac{t^2_{1/2}}{\Delta_{1/2}}\right)(|E^{(1)}_{zx}|^2 - |E^{(2)}_{zx}|^2).
	\end{split}
\end{align}
Note that in Eq.~\eqref{eq: V_alpha sigma_00}, we have assumed $\eta(t)\ll 1$ and omitted all second‑order terms in $\eta^2$. 
Here, $t_{1/2}\equiv\langle J=1/2||e{\bf r}||J'=1/2\rangle$, $t_{3/2}\equiv\langle J=1/2||e{\bf r}||J'=3/2\rangle$ are the reduced matrix elements, with $t_{3/2}\approx\sqrt{2}t_{1/2}$.
For the chosen wavelength $\lambda_0=768$\,nm, we have $\Delta_{3/2}=-h\times 661.5$\,GHz and $\Delta_{1/2}=h\times 1068.6$\,GHz.
This corresponds to an overall red detuning; therefore $V_0,V_z>0$ provided $|E^{(1)}_{zx}|>|E^{(2)}_{zx}|$.

As shown in Fig.~\ref{fig1}(b), Raman couplings between $\ket{g,\sigma}$ and $\ket{e,\bar{\sigma}}$ ($\bar{\sigma}$ denotes the opposite spin) are driven by two‑photon processes involving the fields $\mathbf{E}_1$ and $\mathbf{E}_3$. The corresponding Raman potentials read
\begin{align}
	\begin{split}
	{\cal M}_{g\uparrow, e\downarrow} =& \sum_{J,F}\left( \frac{\Omega^{(J)}_{g\uparrow,F;3,0} \Omega^{(J)*}_{e\downarrow,F;1,+}}{\Delta_{J}}+ \frac{\Omega^{(J)}_{g\uparrow,F;3,-}\Omega^{(J)*}_{e\downarrow,F;1,0}}{\Delta_{J}} \right), \\
	{\cal M}_{g\downarrow, e\uparrow} =& \sum_{J,F}\left( \frac{\Omega^{(J)}_{g\downarrow,F;3,0} \Omega^{(J)*}_{e\uparrow,F;1,-}}{\Delta_{J}}+ \frac{\Omega^{(J)}_{g\downarrow,F;3,+}\Omega^{(J)*}_{e\uparrow,F;1,0}}{\Delta_{J}} \right). \\
	\end{split}
\end{align}
After evaluation we obtain
\begin{align}	
	\begin{split}
{\cal M}_{g\uparrow, e\downarrow} =&-\frac{M_1}{\sqrt{2}}e^{\ui \phi'_\beta}\cos(k_0z+\phi_z)\sin(k_0y-\beta_L/2) \\
	&+\ui \frac{M_1}{\sqrt{2}}e^{\ui \phi'_\alpha}\cos(k_0z+\phi_z)\sin(k_0x-\alpha_L/2),\\
	{\cal M}_{g\downarrow, e\uparrow} =&-\frac{M_1}{\sqrt{2}}e^{\ui \phi'_\beta}\cos(k_0z+\phi_z)\sin(k_0y-\beta_L/2)\\
	&-\ui \frac{M_1}{\sqrt{2}}e^{\ui \phi'_\alpha}\cos(k_0z+\phi_z)\sin(k_0x-\alpha_L/2),
	\end{split}
\end{align}
with the definitions $\phi'_\alpha\equiv\phi'_z- \alpha' - \alpha_L/2 -\pi/2$, $\phi'_\beta\equiv\phi'_z- \beta'-\beta_L/2 -\pi/2$, and
\begin{align}
M_1 = \frac{\sqrt{10}}{27} \left(\frac{t^2_{3/2}}{\Delta_{3/2}}-\frac{2t^2_{1/2}}{\Delta_{1/2}}\right)|E^{(1)}_{zx}||E_{xz}|.
\end{align} 
Raman couplings between $\ket{g,\sigma}$ and $\ket{e,\sigma}$ are driven by two-photon processes induced by the fields $\mathbf{E}_2$ and $\mathbf{E}_4$, 
which give rise to the Raman potentials
\begin{align}
	{\cal M}_{g\sigma,e\sigma}= \sum_{J,F}\left( \frac{\Omega^{(J)}_{g\sigma,F;4,+} \Omega^{(J)*}_{e\sigma,F;2,+}}{\Delta_{J}} + \frac{\Omega^{(J)}_{g\sigma,F;4,-} \Omega^{(J)*}_{e\sigma,F;2,-}}{\Delta_{J}}\right).
\end{align}
Explicitly,
\begin{align}
	\begin{split}
	{\cal M}_{g\uparrow,e\uparrow} 
	=& -\ui \frac{M_2}{2}  \big[e^{\ui \phi_\alpha }\cos(k_0x-\alpha_L/2)\\
	&+e^{\ui \phi_\beta}\cos(k_0y-\beta_L/2)\big]\sin[k_0z+\eta(t)+\phi_z], \\
	{\cal M}_{g\downarrow,e\downarrow} 
	 =& - \frac{M_2}{2} \big[e^{\ui \phi_\alpha }\cos(k_0x-\alpha_L/2) \\
	& +e^{\ui \phi_\beta}\cos(k_0y-\beta_L/2)\big]\sin[k_0z-\eta(t)+\phi_z] ,
	\end{split}
\end{align}
where $\phi_\alpha\equiv \phi'_z- \alpha- \alpha_L/2$, $\phi_\beta\equiv\phi'_z- \beta - \beta_L/2 $ (note that $\phi_\beta-\phi_\alpha = \delta \phi$), and
\begin{align}
	M_2 = \frac{4\sqrt{5}}{27}\left(\frac{t^2_{3/2}}{\Delta_{3/2}}-\frac{2t^2_{1/2}}{\Delta_{1/2}}\right)|E^{(2)}_{zx}||E_{xz}|.
\end{align}

In the $x$-$y$ plane, the lattice potential forms a checkerboard pattern, with neighboring sites aligned along the diagonal directions $x' = (x-y)/\sqrt{2}$ and $y' = (x + y)/\sqrt{2}$. 
It is therefore convenient to adopt the $(x',y',z)$ coordinate frame. We further set $\alpha-\alpha'=\beta-\beta'=\pi/4+2n\pi$  (with $n\in\mathbb{Z}$), which can be realized using EOMs.
The phase difference $\delta \phi$ is assumed to be locked to $2n\pi$, achieved via a Michelson interferometer~\cite{Wang2021}. Under these conditions, after applying the gauge transformation 
\begin{align}
\ket{e,\uparrow}\rightarrow e^{\ui (\pi/2-\phi_\alpha)}\ket{e,\uparrow},\quad \ket{e,\downarrow}\rightarrow e^{-\ui\phi_\alpha}\ket{e,\downarrow},
\end{align}
the lattice potential becomes
\begin{align}
	{\cal V}_{\alpha\sigma}
	=& -V_0 \left[\cos^2(kx'+\phi_x)+\cos^2(ky'+\phi_y)\right] \nonumber\\
	&\quad- V_z \cos^2(k_0z+\phi_z),
\end{align}
and the Raman potentials are given by
\begin{align}
	\begin{split}
	{\cal M}_{g\uparrow, e\downarrow} 
	 =& {M_1}\sin(kx'+\phi_x)\cos(ky'+\phi_y)\cos(k_0z+\phi_z)  \\
	&+\ui{M_1}\cos(kx'+\phi_x)\sin(ky'+\phi_y)\cos(k_0z+\phi_z), \\
	{\cal M}_{g\downarrow, e\uparrow} 
	=& {M_1}\sin(kx'+\phi_x)\cos(ky'+\phi_y)\cos(k_0z+\phi_z) \\
	&-\ui{M_1}\cos(kx'+\phi_x)\sin(ky'+\phi_y)\cos(k_0z+\phi_z),
	\end{split}
\end{align}
and
\begin{align}
	\begin{split}
	{\cal M}_{g\uparrow, e\uparrow}
	=&\,{M_2}\cos(kx'+\phi_x)\cos(ky'+\phi_y) \\
	&\,\sin(k_0z+\eta(t)+\phi_z), \\
	{\cal M}_{g\downarrow, e\downarrow}=&-{M_2 }\cos(kx'+\phi_x)\cos(ky'+\phi_y) \\
	&\,\sin(k_0z-\eta(t)+\phi_z),
	\end{split}
\end{align}
where $\phi_x\equiv-(\alpha_L-\beta_L)/4$ and $\phi_y\equiv-(\alpha_L+\beta_L)/4$. 
Expressed in the combined spin–orbital basis $\tau_i\otimes\sigma_j$ ($i,j\in\{0,x,y,z\}$), the resulting Hamiltonian takes the form given in Eq.~\eqref{eq:HamLat} of the main text.

\subsection{Tight-binding model}

In the tight-binding limit and considering only the $s$-bands, the system is described by the Hamiltonian $H_{\rm TB}$ after applying the gauge transformation $U = e^{i (kx'+ky'+k_0z)\sum_{\sigma}\ket{e,\sigma}\bra{e,\sigma}}$:
\begin{align}\label{eq:TB_base}
	H_{\rm TB} =&
	\sum_{m,n,l} \Big\{\Big [
	C_{m+1,n,l}^\dagger\big(-t_0 \tau_z\sigma_0+ \ui s_x t_{\rm so}\tau_y\sigma_x\big)C_{m,n,l} \nonumber\\
	&+C_{m,n+1,l}^\dagger\, \big(-t_0\tau_z\sigma_0+ \ui s_y t_{\rm so} \tau_y\sigma_y\big)C_{m,n,l} \nonumber\\
	& +C_{m,n,l+1}^\dagger\big(-t_z\tau_z\sigma_0+ \ui s_z t_{\rm soz}\tau_y\sigma_z\big)C_{m.n.l}+ \mathrm{H.c.}\Big] \nonumber\\
	&+C_{m,n,l}^\dagger \big[m_z(t)\,\tau_z\sigma_0 + \eta(t)\,t_{\rm on}\,\tau_x\sigma_0\big]C_{m,n,l}\Big\},
\end{align}
where $C_{\bf j}\equiv (c_{{\bf j}, g\uparrow}, c_{{\bf j}, g\downarrow},c_{{\bf j}, e\uparrow},c_{{\bf j}, e\downarrow})^{\rm T}$ denotes the vector of annihilation operators for all internal states at site $\mathbf{j}=(m,n,l)$, and $c_{{\bf j},\alpha\sigma}$ ($c_{{\bf j},\alpha\sigma}^\dagger$) annihilates (creates) an atom in the state $|\alpha,\sigma\rangle$ at that site. Here, $(s_x,s_y,s_z)=(-1,1,-1)$. The (pseudo)spin-conserved hopping amplitudes $t_{0}$ and $t_{z}$ are given by
\begin{align}
	\begin{split}
	& t_0 = -\int d^3\mathbf{r} \phi^*_s(\mathbf{r}) \left[\frac{\mathbf k^2}{2m}+\mathcal V_{\rm latt}(\mathbf r)\right] \phi_s(\mathbf{r}- \sqrt{2}a_0\hat{\mathbf{x}}), \\
	& t_z = -\int d^3\mathbf{r} \phi^*_s(\mathbf{r}) \left[\frac{\mathbf k^2}{2m}+\mathcal V_{\rm latt}(\mathbf r)\right] \phi_s(\mathbf{r}- a_0\hat{\mathbf{z}}),
	\end{split}
\end{align}
the (pseudo)spin-flipped hopping amplitudes $t_{\rm so}$ and $t_{\rm soz}$ are
\begin{align}
	\begin{split}
	& t_{\rm so} = \int d^3\mathbf{r} \phi^*_s(\mathbf{r}) \mathcal{M}_{x}(\mathbf r) \phi_s(\mathbf{r}-\sqrt{2} a_0\hat{\mathbf{x}}), \\
	& t_{\rm soz} = \int d^3\mathbf{r} \phi^*_s(\mathbf{r}) \mathcal{M}_{z}(\mathbf r) \phi_s(\mathbf{r}-a_0\hat{\mathbf{z}}), 
	\end{split}
\end{align}
and the on-site (pseudo)spin-flip amplitude is
\begin{align}
	 t_{\text{on}} = \int d^3\mathbf{r} \phi^*_s(\mathbf{r}) \mathcal{M}_{0}(\mathbf r) \phi_s(\mathbf{r}),
\end{align}
where $a_0=\pi/k_0$ and $\phi_s(\mathbf{r})$ denotes the Wannier function. Performing the Fourier transformation then yields the Bloch Hamiltonian given in Eq.~\eqref{eq:Hamq}.

\begin{figure}
	\includegraphics[width=0.4\textwidth]{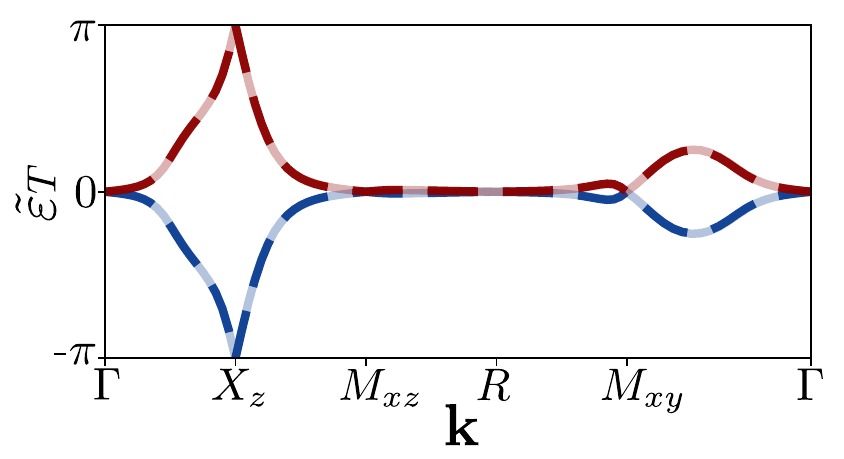}
	\caption{Quasienergy spectra of the Floquet operator computed from the full ORL Hamiltonian (solid line) and the corresponding tight-binding model (dashed line). }
	\label{figS1}
\end{figure}

In the main text, we choose $V_0=2E_r$, $V_z=4E_r$, $M_1=M_2=1E_r$, $m_z^{(1)}=m_z^{(2)}=0.13E_r$, and $\eta_0=0.15$. 
Within the tight-binding description, these correspond to the hopping parameters
$(t_0,t_{z},t_{\rm so},t_{\rm soz},\eta_0t_{\text{on}}) \simeq (0.065,0.097,0.050,0.050,0.10)E_r$. 
We compute the quasienergy spectra of both the full ORL Hamiltonian and the tight‑binding Hamiltonian along the quasimomentum path 
$(k_x,k_y,k_z): \Gamma\,(0,0,0)\rightarrow X_z \,(0,0,\pi)\rightarrow M_{xz}\,(\pi,0,\pi)\rightarrow R\,(\pi,\pi,\pi)\rightarrow M_{xy}\,(\pi,\pi,0)\rightarrow \Gamma\,(0,0,0)$. 
As shown in Fig.~\ref{figS1}, the solid and dashed curves denote the ORL and tight‑binding results, respectively, and are in excellent agreement.

\section{Unpaired Weyl points from the Bloch Hamiltonian}
\label{App:Bloch}

In this Appendix, we determine the positions of the Weyl points from the Bloch Hamiltonian and derive the effective low‑energy Hamiltonian near them.


The Bloch Hamiltonian introduced in the main text can be written as
\begin{equation}
	H(\mathbf{k},t)
	= h_0(\mathbf{k},t) \tau_z\sigma_0
	+ h_4(t) \tau_x \sigma_0+ \tau_y\big[\mathbf{h}_s(\mathbf{k})\cdot\boldsymbol{\sigma}\big],
\end{equation}
where $h_0(\mathbf{k},t) = m_z(t) - 2t_0\sum_{i=x,y}\cos k_i-2t_z\cos k_z$, $h_4(t) = \eta(t)\,t_{\rm on}$, and $\mathbf{h}_s(\mathbf{k}) = (h_1,h_2,h_3) =(-2t_{\rm so}\sin k_x,\; 2t_{\rm so}\sin k_y,\; -2t_{\rm soz}\sin k_z)$.
We define the spin Hamiltonian $H_{s}(\mathbf{k}) = \mathbf{h}_s(\mathbf{k})\cdot\boldsymbol{\sigma}$, which satisfies
\begin{align}
	H_{s}(\mathbf{k}) \ket{ \psi_{\pm}(\mathbf{k})} = \pm s(\mathbf{k}) \ket{\psi_{\pm}(\mathbf{k})},
\end{align}
with $s(\mathbf{k}) = |\mathbf{h}_s(\mathbf{k})| = \sqrt{\sum_{l=1}^3 h^2_l(\mathbf{k})}$. 
Performing the unitary transformation $U(\mathbf{k}) =\id \otimes U^\dagger_s(\mathbf{k})$, 
where $U_s(\mathbf{k}) = \big[|\psi_{+}(\mathbf{k})\rangle\,\, |\psi_{-}(\mathbf{k})\rangle\big]$, we obtain
\begin{align}
	H'(\mathbf{k},t)
	&= U(\mathbf{k})\,H(\mathbf{k},t)\,U^\dagger(\mathbf{k})\nonumber \\
	& = h_0(\mathbf{k},t) \tau_z\sigma_0
	+ h_4(t) \tau_x \sigma_0  + s(\mathbf{k})\tau_y\sigma_z.
\end{align}
In the basis $\{\ket{g,\psi_{+}}, \ket{e,\psi_{+}},\ket{g,\psi_{-}},\ket{e,\psi_{-}}\}$, the transformed Hamiltonian becomes block-diagonal:
\begin{equation}
	H'(\mathbf{k},t)
	= \begin{pmatrix}
		H_{+}(\mathbf{k},t) & 0 \\
		0 & H_{-}(\mathbf{k},t)
	\end{pmatrix},
\end{equation}
with $H_\pm(\mathbf{k},t) = h_0(\mathbf{k},t)\tau_z + h_4(t)\tau_x \pm s(\mathbf{k})\tau_y$. 
Each block satisfies ($n=1,2$)
\begin{align}
	H_{\pm}(\mathbf{k},t) \ket{u^{\pm}_{n}(\mathbf{k},t)} = E_n(\mathbf{k},t) \ket{u^{\pm}_n(\mathbf{k},t)},
\end{align}
with $E_2(\mathbf{k},t)= -E_1(\mathbf{k},t)= \sqrt{\sum_{l=0}^4 h^2_l (\mathbf{k},t)}$. 
Consequently, the two degenerate eigenstates of $H'(\mathbf{k},t)$ with energy $E_1$ are 
$|\Psi^{(1)}(\mathbf{k},t)\rangle=|u^{+}_1(\mathbf{k},t)\rangle\otimes|\psi_{+}(\mathbf{k})\rangle$ and 
$|\Psi^{(2)}(\mathbf{k},t)\rangle= |u^{-}_1(\mathbf{k},t)\rangle\otimes|\psi_{-}(\mathbf{k})\rangle $.
Their time evolution is governed separately by $H_{+}$ and $H_{-}$. 
Over one driving period $T$, the corresponding evolution operators are $U^{(1,2)}_{\mathbf{k}}= \mathcal{T}\exp[-\ui\int_0^T H_{\pm}(\mathbf{k},t) dt]$.
Quasienergies of these two evolution operators become degenerate when the two blocks coincide $H_{+}(\mathbf{k},t)=H_{-}(\mathbf{k},t)$, i.e., $s(\mathbf{k})=0$.
This occurs at the high‑symmetry momenta $\mathbf{k}_j \in  \{ (k_x,k_y,k_z) | k_x,k_y,k_z \in \{0,\pi \} \}$.

In the adiabatic limit, the quasienergies of the lower-energy sector are determined by the Berry phases
\begin{align} \label{eq: Berry}
	\gamma_{\pm}(\mathbf{k}) = \ui \int_0^T dt \bra{u^{\pm}_1 (\mathbf{k},t)} \frac{\partial}{\partial t} \ket{u^{\pm}_1(\mathbf{k},t)},
\end{align}
and the time-evolution operator in this sector can be written as
\begin{align}
{U}_\mathbf{k}^{(-)}=&\,e^{i\gamma_{+}(\mathbf{k})}|\Psi^{(1)}(\mathbf{k},t=0)\rangle\langle\Psi^{(1)}(\mathbf{k},t=0)| \nonumber\\
&+e^{i\gamma_{-}(\mathbf{k})}|\Psi^{(2)}(\mathbf{k},t=0)\rangle\langle\Psi^{(2)}(\mathbf{k},t=0)|.
\end{align}
Treating $s(\mathbf{k})\tau_y$ as a perturbation near a degeneracy point $\mathbf{k}_j$, the eigenstates $\ket{u^{\pm}_{1}(\mathbf{k},t)}$ are approximated to first order as
\begin{align}
	\ket{u^{\pm}_{1}(\mathbf{k},t)} \approx &\ket{u_1(\mathbf{k},t)}\mp  \frac{\bra{u_2(\mathbf{k},t)}\tau_y\ket{u_1(\mathbf{k},t)}}{2E_0(\mathbf{k},t)}\nonumber\\
	&\times s(\mathbf{k})\ket{u_2(\mathbf{k},t)}.
\end{align}
Here, $\ket{u_{1,2}(\mathbf{k},t)}$ are the eigenstates of the Hamiltonian $H_{\rm orb}(\mathbf{k},t) = h_0(\mathbf{k},t)\tau_z + h_4(t)\tau_x$ with eigenvalues
$\mp E_0(\mathbf{k},t)$, where $E_0(\mathbf{k},t)\equiv\sqrt{h^2_0(\mathbf{k},t)+h^2_4(t)}$.
Substituting this expansion into Eq.~\eqref{eq: Berry} gives
\begin{equation}
	\gamma_{\pm}(\mathbf{k}) = \pm  s(\mathbf{k}) \delta \gamma(\mathbf{k}) + \gamma_0(\mathbf{k}),
\end{equation}
with 
\begin{align}
	\begin{split}
	\delta \gamma(\mathbf{k}) &= \int_0^T dt \frac{h_4(t) \partial_t h_0(\mathbf{k},t) - h_0(\mathbf{k},t) \partial_t h_4(t)}{2E^3_0(\mathbf{k},t)}, \\
	\gamma_0(\mathbf{k}) &= \int_0^T dt \frac{h_4(t) \partial_t h_0(\mathbf{k},t) - h_0(\mathbf{k},t) \partial_t h_4(t)}{2E^2_0(\mathbf{k},t)}.
	\end{split}
\end{align}
Projecting  ${U}_\mathbf{k}^{(-)}$ onto the spin subspace (by tracing out the orbital part) yields the reduced Floquet operator
\begin{align}
	\tilde{U}_\mathbf{k} &= e^{i\gamma_0(\mathbf{k})}\left\{\cos[\delta \gamma(\mathbf{k}) s(\mathbf{k})] + \ui\sin[\delta \gamma(\mathbf{k}) s(\mathbf{k})]\frac{H_{s}(\mathbf{k})}{s(\mathbf{k})} \right\} \nonumber\\
	& = e^{i\gamma_0(\mathbf{k})} e^{ \ui \delta \gamma(\mathbf{k}) H_{s}(\mathbf{k}) }.
\end{align}
The Floquet Hamiltonian of the low‑energy sector is defined via $\tilde{U}_\mathbf{k}=e^{-\ui \tilde{H}_F({\bf k})T}$.
From the expression above it follows directly that $ \tilde{H}_F({\bf k})$ is governed by the spin Hamiltonian $H_{s}(\mathbf{k})$. 
Expanding near a degeneracy point $\mathbf{k}_j$ produces the Weyl form presented in Eq.~\eqref{eq: HF_low energy} of the main text.

\begin{widetext}

\section{Effective Hamiltonian with a synthetic magnetic field}\label{App:Ham_with_B}

In this section, we derive the effective Hamiltonian presented in Eq.~\eqref{eq:Heff_Azx} of the main text. We start with the tight‑binding Hamiltonian
\begin{align}
	H_{\rm TB} =&
	\sum_{m,n,l} \Big\{\Big [
	C_{m+1,n,l}^\dagger\big(-t_0 \tau_z\sigma_0+ \ui s_x t_{\rm so}\tau_y\sigma_x\big)C_{m,n,l} 
	+C_{m,n+1,l}^\dagger\big(-t_0\tau_z\sigma_0+ \ui s_y t_{\rm so} \tau_y\sigma_y\big)C_{m,n,l}+C_{m,n,l+1}^\dagger \nonumber\\
	& \big(-t_z\tau_z\sigma_0+ \ui s_z t_{\rm soz}\tau_y\sigma_z\big)C_{m,n,l}+ \mathrm{H.c.}\Big]+C_{m,n,l}^\dagger 
	 \big[m_z(t)\tau_z\sigma_0 + \eta(t)t_{\rm on}\tau_x\sigma_0+V_{m,n,l}(t)\big]C_{m,n,l}\Big\},
\end{align}
where $V_{m,n,l}(t)= n\Delta + \frac{V_K}{2}\cos\!\big(\omega t+ \delta \mathbf{k \cdot R}\big)$, 
with $\mathbf{R} =  a_0(\sqrt{2}m\hat{\mathbf x}+ \sqrt{2} n\hat{\mathbf y}+ l\hat{\mathbf z})$ 
denoting the position of the lattice site $(m,n,l)$.
We then apply the unitary transformation
\begin{equation}
	U(t)=\exp\!\Big[\ui\sum_{m,n,l}\Lambda_{m,n,l}(t)\,C_{m,n,l}^\dagger C_{m,n,l}\Big],
\end{equation}
with $\Lambda_{m,n,l}(t) = n\Delta t+\frac{V_K}{2\omega}\sin\!\big(\omega t+\delta \mathbf{k \cdot R}\big)$. This yields the transformed Hamiltonian
\begin{align}\label{eq:TB_base_With_B}
	H'_{\rm TB} =&
	\sum_{m,n,l} \Big\{\Big [
	C_{m+1,n,l}^\dagger e^{\ui \theta_{m}(t)}\big(-t_0 \tau_z\sigma_0+ \ui s_x t_{\rm so}\tau_y\sigma_x\big)C_{m,n,l}+C_{m,n+1,l}^\dagger e^{\ui \theta_{n}(t)}\big(-t_0\tau_z\sigma_0+ \ui s_y t_{\rm so} \tau_y\sigma_y\big)C_{m,n,l}  \nonumber\\
	& +C_{m,n,l+1}^\dagger e^{\ui \theta_{l}(t)}\big(-t_z\tau_z\sigma_0+ \ui s_z t_{\rm soz}\tau_y\sigma_z\big) C_{m.n.l} + \mathrm{H.c.}\Big]
	+C_{m,n,l}^\dagger \big[m_z(t) \tau_z\sigma_0 + \eta(t) t_{\rm on} \tau_x\sigma_0\big]C_{m,n,l}\Big\},
\end{align}
where
\begin{align}
	\begin{split}
	&\theta_m(t) = \frac{V_K}{\omega} \cos(\omega t +\delta \mathbf{k\cdot R} +\frac{\delta \mathbf{k}\cdot  a_0\hat{\mathbf{x}}}{\sqrt{2}} )
	\sin(\frac{\delta \mathbf{k}\cdot a_0\hat{\mathbf{x}}}{\sqrt{2}}),\\
	&\theta_n(t)  = \Delta t + \frac{V_K}{\omega} \cos(\omega t +\delta \mathbf{k\cdot R} +\frac{\delta \mathbf{k}\cdot  a_0\hat{\mathbf{y}}}{\sqrt{2}} )
	\sin(\frac{\delta \mathbf{k}\cdot a_0\hat{\mathbf{y}}}{\sqrt{2}}),\\
	&\theta_l(t) = \frac{V_K}{\omega} \cos(\omega t +\delta \mathbf{k\cdot R} +\frac{\delta \mathbf{k}\cdot  a_0\hat{\mathbf{z}}}{{2}} )
	\sin(\frac{\delta \mathbf{k}\cdot a_0\hat{\mathbf{z}}}{{2}}).
	\end{split}
\end{align}

Using the Jacobi–Anger expansion $e^{\ui z \cos\theta} = \sum_{n=-\infty}^\infty \ui^n \mathcal{J}_n(z) e^{\ui n \theta}$, 
where $\mathcal{J}_n$ is the $n$-th order Bessel function of the first kind, and working in the high‑frequency resonance limit $\Delta = \omega \gg t_0$, 
we obtain the effective Hamiltonian
\begin{align}
	H_{\rm eff}=&
	\sum_{m,n,l} \Big\{\Big [
	C_{m+1,n,l}^\dagger\big(-t^x_0 \tau_z\sigma_0+ \ui s_x t_{\rm so}^{x}\tau_y\sigma_x\big)C_{m,n,l}
	+C_{m,n+1,l}^\dagger\, e^{-\ui \phi_{m,n,l}}\big(-t^y_0\tau_z\sigma_0+ \ui s_y t_{\rm so}^{y} \tau_y\sigma_y\big)C_{m,n,l} \nonumber\\
	& +C_{m,n,l+1}^\dagger\big(-t^z_z\tau_z\sigma_0+ \ui s_z t^z_{\rm soz}\tau_y\sigma_z\big)C_{m,n,l}+ \mathrm{H.c.}\Big] 
	+C_{m,n,l}^\dagger \big[m_z(t)\,\tau_z\sigma_0 + \eta(t)\,t_{\rm on}\,\tau_x\sigma_0\big]C_{m,n,l}\Big\},
\end{align}
where the site‑dependent Peierls phase is
\begin{align}
	\phi_{m,n,l} &= \delta \mathbf{k \cdot  R}+\frac{\delta \mathbf{k}\cdot a_0\hat{\mathbf{y}}}{\sqrt{2}}-\frac{\pi}{2},
\end{align}
and the renormalized hopping amplitudes read
\begin{align}
	t_{0,\rm so}^{x} = t_{0,\rm so}\, \mathcal{J}_{0}\!\!\left[\frac{V_K}{\Delta}\sin(\frac{\delta \mathbf{k}\cdot a_0\hat{\mathbf{x}}}{\sqrt{2}})\right],\quad
	t_{0,\rm so}^{y} = t_{0,\rm so}\,\mathcal{J}_{1}\!\!\left[\frac{V_K}{\Delta}\sin(\frac{\delta \mathbf{k}\cdot a_0\hat{\mathbf{y}}}{\sqrt{2}})\right],  \quad
	t^z_{z,\rm soz} = t_{z,\rm soz}\, \mathcal{J}_{0}\!\!\left[\frac{V_K}{\Delta}\sin(\frac{\delta \mathbf{k}\cdot a_0\hat{\mathbf{z}}}{{2}})\right].
\end{align}
In the main text, we specialize to $\delta \mathbf{k}\cdot a_0\hat{\mathbf{x}} = \sqrt{2}\pi\beta\sin\theta$, $\delta \mathbf{k}\cdot a_0\hat{\mathbf{y}} = \sqrt{2}\pi\beta\cos\theta$, and $\delta \mathbf{k}\cdot a_0\hat{\mathbf{z}} = 0$.

\end{widetext}

\appendix


\begin{thebibliography}{99}
\bibitem{footnote} The lifetime $\tau$ of ultracold $^{40}$K atomic systems in a 3D ORL is shorter than that in a 2D ORL. 
However, the adiabatic condition $T\geq500 E^{-1}_r$ adopted in the main text is relatively strict; in principle, it can be relaxed to $T=200$--$300E^{-1}_r$ 
while still producing fairly good near‑adiabatic results [see Fig.~\ref{fig4}(b)]. Therefore, in experiments with 3D ORLs, it remains possible to find a parameter region where $T<\tau$.
\end{thebibliography}
\end{document}